\documentclass[11pt]{article}

\usepackage{mathtools}
\usepackage{amssymb}
\usepackage{dsfont}
\usepackage{slashed}
\usepackage{fullpage}
\usepackage{siunitx}
\usepackage{ragged2e,graphicx,subcaption}
\usepackage{diagbox}
\usepackage{titlesec}
\titleformat*{\section}{\normalsize\bfseries}
\titleformat*{\subsection}{\normalsize\bfseries}
\titleformat*{\subsubsection}{\normalsize\bfseries}
\usepackage{cite}
\usepackage{amsmath,amssymb,braket,tikz}
\usetikzlibrary{tikzmark,calc}
\usepackage[colorlinks=true,linkcolor=blue,citecolor=magenta,linktocpage=true]{hyperref}

\usepackage[english]{babel}
\usepackage{csquotes}
\MakeOuterQuote{"}

\DeclareMathAlphabet{\bbvar}{U}{BOONDOX-ds}{m}{n}

\makeatletter
\renewcommand{\@dotsep}{10000}
\makeatother

\usepackage[makeroom]{cancel}
\usepackage{ulem} 

\usepackage{lipsum}
\usepackage{mathtools}
\usepackage{xcolor}
\usepackage[colorlinks=true,linkcolor=blue,citecolor=magenta,linktocpage=true]{hyperref}
\usepackage{physics}
\usepackage{csquotes}
\usepackage{interval}
\usepackage{bm}

\usepackage{caption}
\usepackage{subcaption}

\usepackage[makeroom]{cancel}

\def\be{\begin{equation}}
	\def\ee{\end{equation}}
\def\bea{\begin{eqnarray}}
	\def\eea{\end{eqnarray}}
\def\beq{\begin{eqnarray}}
	\def\eeq{\end{eqnarray}}
\def\bas{\begin{subequations}\begin{eqnarray}}
		\def\eas{\end{eqnarray}\end{subequations}}
\def\nn{\nonumber}

\newcommand{\cR}{{\mathcal R}}
\newcommand{\cO}{{\mathcal O}}

\newcommand{\cV}{{\mathcal V}}

\def\rd{\textrm{d}}

\newcommand{\bes}{\begin{eqnarray}}
	\newcommand{\ees}{\end{eqnarray}}

\def\dd{\mathrm{d}}

\def\nn{\nonumber}

\def\nn{\nonumber}

\numberwithin{equation}{section}

\begin{document}

\title{
\Large{\textbf{\sffamily Black hole photon ring beyond General Relativity: \\ an integrable parametrization}}
}
\author{\sffamily Jibril Ben Achour\;$^{1,2,3}$,  \'Eric Gourgoulhon\;$^{4,5}$, Hugo Roussille\;$^{3}$}
\date{\small{\textit{$^{1}$ Arnold Sommerfeld Center for Theoretical Physics, Munich, Germany, \\
$^{2}$ Munich Center for Quantum Sciences and Technology, Germany,\\
$^{3}$ Universit\'e de Lyon, ENS de Lyon, Laboratoire de Physique, CNRS UMR 5672, Lyon 69007, France,\\
 $^{4}$ LUX, CNRS UMR 8262, Observatoire de Paris - PSL, Sorbonne Universit\'e Paris, 5 place Jules Janssen, 91190 Meudon, France \\
 $^{5}$ Laboratoire de Math\'ematiques de Bretagne Atlantique, CNRS UMR 6205, Universit\'e de Bretagne Occidental, 6 avenue Victor Le Gorgeu, 29200 Brest, France }}}

\maketitle

\begin{abstract}
In recent years, the shape of the photon ring in black holes images has been argued to provide a sharp test of the Kerr hypothesis for future black hole imaging missions. In this work, we confront this proposal to beyond Kerr geometries and investigate the degeneracy in the estimations of the black hole parameters using the circlipse shape proposed by Gralla and Lupsasca. To that end, we consider a model-independent parametrization of the deviations to the Kerr black hole geometry, dubbed Kerr off shell (KOS),  which preserves the fundamental symmetry structure of Kerr known as the Killing tower. Besides exhibiting a Killing tensor and thus a Carter-like constant, all the representants of this family also possess a Killing-Yano tensor and are of Petrov type D. The allowed deviations to Kerr, selected by the symmetry, are encoded in two free functions which depend respectively on the radial and polar angle coordinates. Using the symmetries, we provide an analytic study of the radial and polar motion of photon trajectories generating the critical curve, to which the subrings composing the photon ring converge. This allows us to derive a ready-to-use closed formula for the parametric critical curve in term of the free functions parametrizing the deviations to Kerr. Using this result, we confront the circlipse fitting function to four examples of Kerr-like objects and we show that it admits a high degree of degeneracy. At a given inclination, the same circlipse can fit both a Kerr black hole of a given mass and spin $(M,a)$ or a modified rotating black hole with different mass and spin parameters $(M,a)$ and a new parameter $\alpha$. Therefore, future tests of the Kerr hypothesis could be achieved only provided one can measure independently the mass and spin of the black hole to break this degeneracy.
\end{abstract}

\thispagestyle{empty}
\newpage
\setcounter{page}{1}

\hrule
\tableofcontents
\vspace{0.7cm}
\hrule

\newpage


The first interferometric observations of the supermassive black hole at the center of the M87 galaxy by the Event Horizon Telescope (EHT) in 2019 \cite{EventHorizonTelescope:2019dse, EventHorizonTelescope:2019ths, EventHorizonTelescope:2019pgp}, followed by the first image of the Sagittarius A$^{\ast}$ black hole at the center of our galaxy in 2022 \cite{EventHorizonTelescope:2022wkp, EventHorizonTelescope:2022wok}, has opened a new era for horizon scale electromagnetic astronomy and a new window to test the strong field regime of gravity. These observations have revealed the existence of an unresolved and asymmetric bright ring, as expected from the theoretical predictions and simulations. Since then, important efforts have been devoted to clarify how these interferometric signatures of ultra-compact objects could be used to test General Relativity (GR) and its different extensions\footnote{Important efforts have focused on investigating the shape of the photon ring of exotic compact objects, among which hairy black holes \cite{Cunha:2015yba, Moffat:2015kva, Cunha:2016bpi, Vincent:2016sjq, Cunha:2018acu, Cunha:2019ikd, Khodadi:2020jij, BenAchour:2020fgy, Long:2020wqj, Gan:2021xdl, Kumar:2021cyl, Afrin:2021imp, Antoniou:2022dre, Wang:2023vcv, Rodriguez:2024ijx}, wormholes \cite{Ohgami:2015nra, Shaikh:2018kfv,Lamy:2018zvj,Rahaman:2021web}, boson stars \cite{Vincent:2015xta,Grandclement:2016eng}, parity-violating black holes~\cite{Chen:2020aix}, regular black holes~\cite{Tsukamoto:2017fxq} and quantum corrected black holes \cite{Shu:2024tut}. To a large extent, these study use numerical ray-tracing methods to determine the shadow of the ultra-compact object. Theory-agnostic results relating the size of black hole hair to the size of the photon ring have been presented in~\cite{Ghosh:2023kge, Ghosh:2025igz}.} \cite{Tsukamoto:2014tja, Khodadi:2020gns, Gralla:2020pra, Vincent:2020dij, Khodadi:2021gbc, Glampedakis:2021oie, Stepanian:2021vvk, Vagnozzi:2022moj, Broderick:2023jfl, Staelens:2023jgr, Khodadi:2024ubi, Yue:2025fly, Carballo-Rubio:2025zwz }.

The gravitational lensing induced by a Kerr black hole\footnote{See \cite{Wang:2022ouq, Chen:2024oyv} for the study of the lensing induced by a Kerr-Newman black hole.} on photons emitted from a generic source possesses universal properties predicted by GR. In particular, photons can be momentarily captured and loop arbitrary many times around the black hole before escaping to infinity. These photons generically produce on the screen detector an infinite sequence of photon subrings corresponding to the arbitrary number of loops made around the hole \cite{Gralla:2019xty}. These subrings converge towards a curve called the \textit{critical curve} which stands as a key prediction of GR first derived in  \cite{Bardeen}. In this work, the \textit{photon ring} will refer to the limit of the photon subrings and therefore be confounded with the critical curve\footnote{In the recent literature, the term \enquote{photon ring} is rather used for the set of all photon subrings~\cite{Johnson:2019ljv,Paugnat:2022qzy}.}. It depends on the black hole mass $M$, spin $a$ and the observer inclination $y_{\cO}$ (w.r.t the black hole rotation axis). This critical curve, being the limit of the sequence of photon subrings, is not an observable. Nevertheless, as the sequence converges quickly, the critical curve is the fundamental theoretical object upon which an observable can be built (see Ref.~\cite{Paugnat:2022qzy} for a discussion and limitations). In particular, it is expected to be contained in the the so far unresolved bright ring observed by the EHT. This challenges the possibility to directly test the shape of the photon ring, and thus the GR prediction, with current data. Observation of the primary photon ring and subsequent subrings will require a space-based interferometric mission \cite{Johnson:2024ttr,Paugnat:2022qzy}. 

In the last five years, important efforts have been done to i) characterize universal observables free from astrophysical bias allowing for a sharp test of GR \cite{Gralla:2017ufe, Johnson:2019ljv, Himwich:2020msm, Gralla:2020nwp, Gralla:2020srx, Hadar:2022xag, Cardenas-Avendano:2023dzo, Jia:2024mlb}, and ii) to clarify the feasibility of a future space-based mission aiming at measuring these observables \cite{Gralla:2020yvo, Paugnat:2022qzy, Vincent:2022fwj, Salehi:2024cim, KumarWalia:2024omf}. In a series of works, it was shown how the intensity profile of the photon (sub)ring image (the amplitude and phase of the signal) depends on its geometrical shape (the diameter and width of the ring) \cite{Gralla:2020nwp, Gralla:2020srx}. This relation allows  one to reconstruct the shape of the photon ring starting from possibly partial information, for instance if one has only access to the amplitude of the signal. Prospects for the measurement of this universal signatures using a space-based interferometric  mission was further explored with encouraging results, motivating the recent Black Hole Explorer (BHEX) proposal \cite{Johnson:2024ttr, Lupsasca:2024xhq}. With these exciting recent developments allowing to test GR and its extensions using photon ring science in the near future, it becomes crucial to provide tractable ready-to-use parametrization of beyond Kerr geometries. The key goal is to provide a sufficiently general parametrization encompassing a large set of possible and physically reasonable deviations, while at the same time allowing one to analytically derive the photon ring observables mentioned above, i.e. the critical curve and its intensity profile. 

To that end, it is crucial that such new parametrization preserves the explicit and hidden symmetries of the Kerr geometry which are responsible for the integrability of the geodesic motion, and thus to obtain close analytical expression for the critical curve. A first initial agnostic parametrization allowing for a Carter-like constant was presented by Johanssen and Psaltis (JP) in \cite{Johannsen:2013szh, Johannsen:2013vgc}, followed by a series of works generalizing the JP proposal \cite{Cardoso:2014rha, Papadopoulos:2018nvd, Konoplya:2018arm, Carson:2020dez, Carson:2020iik, Yagi:2023eap, Salehi:2023eqy}. So far all these new proposals have focused on including deformation to the radial potential experienced by test particles while still preserving the existence of a non-trivial rank-$2$ Killing tensor which generates a Carter-like constant. However, this is not the most fundamental hidden symmetry of the Kerr black hole. Indeed, the Carter charge and the associated Killing tensor of the Kerr geometry, respectively identified by Carter in \cite{Carter:1968ks} and Walker and Penrose in \cite{WalkeP70}, actually descend from a hierarchy of Killing objects known as the Killing tower. This structure was first identified  in the context of higher dimensional rotating black holes in \cite{Krtous:2006qy, Kubiznak:2006kt, Krtous:2008tb}. Concretely, the Killing tower is built from the existence of a non-degenerate conformal Killing-Yano rank-2 tensor from which all other Killing objects can be canonically derived. Interestingly, the family of geometries which admits such Killing tower (and not only the existence of its Killing tensor) has been fully characterized and is dubbed the \textit{Kerr off shell} (KOS) family (see \cite{Frolov:2017kze} for a review). The associated metric is given by (\ref{offshell}) below. The goal of this work is to use this new KOS family of geometries as a new agnostic parametrization for Kerr-like objects and show the new interesting features it allows. There are two main outcomes coming from this new parametrization. 

 First, the KOS family is parametrized by two free functions denoted hereafter $(\Delta_r(r), \Delta_y(y))$ where $y \propto \cos{\theta}$ refers to the polar angle. These free functions modify respectively the radial and polar potentials (\ref{geopot}) experienced by test particles. Comparing to the previous parametrization \cite{Papadopoulos:2018nvd, Konoplya:2018arm, Carson:2020dez, Carson:2020iik, Yagi:2023eap}, the free function $\Delta_r(r)$ encompassed a large set of radial deformations treated so far and coincides with the deformation introduced in \cite{Yagi:2023eap} (denoted $A_5(r)$ there). However, to our knowledge, the polar deformation encoded in $\Delta_y(y)$ is new. Therefore, the KOS parametrization allows for a new deformation of the Kerr geometry while still preserving its most fundamental hidden symmetries. As we shall see, the new polar deformation allows to parametrize new subtle modifications affecting the critical angles $\theta_{\pm}$ characterizing the polar motion of the test particles which plays a non-trivial role when computing the image of source lensed by a Kerr-like object. Notice that general polar deformations to Kerr have already been considered in \cite{Konoplya:2016jvv}, but the parametrization introduced in that work does not preserve the rank-2 Killing tensor.

Second, since the KOS family is defined by the Killing tower structure, the geodesic motion is integrable and one can derive a generalized Carter constant given by (\ref{Carter}). Quite remarkably, the form of the KOS metric and the simple expression for its Killing objects allows one to obtain the closed formula (\ref{eq:coord-critical-curve}) describing the parametric shape of the critical curve in terms of the free functions $(\Delta_r(r), \Delta_y(y))$. This close formula is the main result of this work. It provides a ready-to-use parametric expression for the critical curve which allows us to test any given deformations encompassed in the KOS family. 

In the second part of this work, we use this new general parametric expression for the KOS critical curve to explore the robustness of testing GR using the shape of the photon ring as proposed by Gralla and Lupsasca in \cite{Gralla:2020srx}. The strategy adopted in \cite{ Gralla:2020srx, Gralla:2020yvo} relies on fitting the parametric expression for the critical curve by a \textit{phoval curve} (or equivalently a circlipse) on the observer's screen. The even contribution to this fitting curve encoding the projected diameter $d_{\varphi}$, is characterized by three real parameters $(R_0, R_1, R_2)$ while its odd contribution, i.e. the centroid $C_{\varphi}$, involves two additional parameters. The idea is that focusing on the even contribution $d_{\varphi}$, one can fit to a great accuracy the Kerr critical curve and relate the three parameters $(R_0, R_1, R_2)$ to the black hole mass and spin $(M,a)$ and to the observer inclination $y_{\cO}$. In turn, the amplitude of the visibility (the Fourier transform) of the photon ring image is uniquely related to the profile $d_{\varphi}$ \cite{Gralla:2020nwp}. While this suggests that the measurement of the shape of the photon ring can provide a sharp test of the Kerr hypothesis, and thus of GR, the procedure which consists in introducing a fitting \textit{phoval} function could a priori remains blind to other Kerr-like geometries and thus admits some degeneracy in the parameter estimations. To understand these possible degeneracies, we confront the \textit{phoval fitting function} proposed in \cite{Gralla:2020yvo} to four examples of Kerr-like geometries which involve a new parameter (additionally to the mass and spin) and which belong to our KOS family. Besides two well known examples given by the Kerr-MOG geometry and the regular Simpson-Visser rotating black hole, we consider two new modifications of Kerr which have not been treated so far and which are allowed by the KOS parametrization. For each of the four examples, we present the following results:
\begin{itemize}
\item First, we provide analytical expressions for the radial and polar extend of the photon ring and for the angular momentum $\ell_c$ and Carter constant $k_c$ labeling critical spherical photon orbits. Notice that these quantities are usually not accessible due to either the high complexity of the model or to its non-integrability. 
\item Second, for each examples, we test the degeneracy in the parameter estimation  using the circlipse fitting function. We show that at maximal inclination, i.e. $y_{\cO} =0 $ which corresponds to the equatorial plane, an accurate fit of the Kerr black hole for a given $(M, a)$ can always be degenerate with another equally accurate fit for the Kerr-like object for another set of parameters $(M, a, \alpha)$ where $\alpha$ is the new parameter encoding the deviation to Kerr. Finally, we provide a plot of the shadow for each examples for different values of the new parameter $\alpha$.
\end{itemize}

This article is organized as follows. In Section~\ref{KOS}, we present the \textit{Kerr off shell} (KOS) family involving the two free functions and we review the Killing tower structure. We further compare this new parametrization of Kerr-like objects with previous works. Section~\ref{geod} is devoted to the geodesic motion in the generic KOS geometry. We derive the Killing tensor charge and discuss the radial and polar motion and their associated potentials (\ref{potgeo1} - \ref{potgeo2}). Finally, we present a detail discussion on the photon ring structure in terms of the two free functions. In Section~\ref{sec:shadow}, we derive the parametric form of the critical curve, given in (\ref{eq:coord-critical-curve}), for the KOS family. The last Section~\ref{sec:examples} is devoted to the study of several concrete examples of beyond Kerr geometries which are used to test the limitations of the \textit{phoval fitting function}. We conclude by a discussion of our results and the perspectives it opens.

\section{The Kerr off shell family}
	
	\label{KOS}
	
	In this section, we introduce the new parametrization for Kerr-like objects which we will use throughout this work: the Kerr off-shell family of geometries.  We present briefly its ruling structure, namely the Killing tower and we compare this new beyond-Kerr-parametrization to the existing ones.

	\subsection{Preserving Kerr hidden symmetries}
	
By virtue of the no-hair theorem of four dimensional GR
(see \cite{Chrusciel:2012jk} for a review), the Kerr geometry is the unique asymptotically flat stationary and axi-symmetric vacuum solution of GR. Therefore, it stands as our best model for the observed rotating ultracompact astrophysical objects, in particular supermassive black holes sitting at the center of galaxies. Nevertheless, even if no observational evidence point towards any deviation to this model so far, it is crucial to provide efficient parametrization of possible deviations that could be captured in the future.

As is well known, the Kerr geometry enjoys a set of explicit and hidden symmetries which play a key role in the derivations of its properties. Besides the time and azimutal isometries, it also admits a Killing-Yano two-form which canonically equips the geometry with a Killing tensor. This later object generates the well-known Carter constant which makes the geodesic motion integrable \cite{Carter:1968ks, Walker:1970un, Carter:1977pq}, allowing us to classify the different allowed trajectories of massive and massless particles on the Kerr geometry \cite{Gralla:2019ceu, Compere:2021bkk}. In turn, the presence of a Killing-Yano tensor generates a set of conserved charges which allows one to integrate the motion of any test spinning particle\footnote{A set of conserved charges for an extended quadrupolar test object on the Kerr background was found in \cite{Compere:2021kjz, Compere:2023alp} assuming that the test object is a black hole. It was further shown in \cite{Ramond:2024ozy} that a sufficient number of these conserved charges are in involution, ensuring the integrability of this system.} (up to linear order in the spin) on the Kerr background \cite{Ramond:2022vhj, Ramond:2024ozy}, allowing us to study properties of extreme-mass-ratio inspirals \cite{Witzany:2019nml, Skoupy:2024uan}. Finally, the same hidden Killing symmetries are also responsible for the well-known separability of spin-s wave equations allowing one to study the scattering problem of various test fields with the Kerr black hole \cite{Teukolsky:1972my}. Therefore, testing the Kerr hypothesis can be understood as testing the fundamental symmetries of the Kerr black hole. From this point of view, it appears interesting to search for a parametrization of Kerr-like geometries which preserves the Kerr symmetries. 

Consider the following stationary and axi-symmetric metric
	\begin{align}
		\label{offshell}
		\dd s^2 = - \frac{\Delta_r}{\Sigma} \left( \dd \tau + y^2 \dd \varphi \right)^2 + \frac{\Delta_y}{\Sigma} \left( \dd \tau - r^2 \dd \varphi \right)^2 + \frac{\Sigma}{\Delta_r} \dd r^2 + \frac{\Sigma}{\Delta_y} \dd y^2
	\end{align}
	where
	\begin{equation}
		\Sigma = r^2 + y^2 \qquad \Delta_r:= \Delta_r(r) \qquad \Delta_y:= \Delta_y(y)
	\end{equation}
	while the inverse metric reads
	\begin{align}
		g^{\mu\nu} \partial_\mu\partial_\nu = \frac{1}{\Sigma} \left[ - \frac{1}{\Delta_r} \left( r^2 \partial_\tau + \partial_\varphi \right)^2 + \frac{1}{\Delta_y} \left( y^2 \partial_\tau -  \partial_\varphi \right)^2 + \Delta_r (\partial_r)^2 + \Delta_y (\partial_y)^2 \right]
	\end{align} 
	The metric is dubbed \textit{off shell} because it does not have to be a solution of Einstein's equations: it is parametrized by the two free functions $\Delta_r(r)$ and $\Delta_y(y)$. The metric (\ref{offshell}) is a subset of 
	a general class of metrics considered by Carter
	\cite{Carter:1968ks} to ensure the separability of the 
	Hamilton-Jacobi and Schrödinger equations. In this subset, the function $\Sigma$ is not modified w.r.t the Kerr case. Contrary to the geometries of the general class considered by Carter which are not algebraically special, this restriction ensures that the spacetime geometry is of Petrov type D.
	
	While the functions $\Delta_r$ and $\Delta_y$ are free, they must satisfy several constraints in order for the spacetime~\eqref{offshell} to describe our object of interest, i.e. an asymptotically flat black hole. First, asymptotic flatness constrains the asymptotic behavior of $\Delta_r$: one must have $\Delta_r \sim r^2$ as $r \to \infty$. Second, in order for the metric~\eqref{offshell} to have an event horizon, $\Delta_r$ must be such that $\Delta_r(r) > 0$ for $r > r_\mathrm{h}$ and $\Delta_r(r_\mathrm{h}) = 0$; $r_h$ corresponds to the radius of the outermost horizon. Third, $\Delta_y$ must be positive for all possible values of $y$ to conserve the Lorentzian signature. Finally, since the metric exhibits a curvature singularity for $\Sigma = 0$, we assume $r_\mathrm{h} > 0$ in order to avoid the presence of a naked singularity. Furthermore, a metric satisfying these constraints might still exhibit a conical or Misner string singularity, depending on the exact shape of the functions $\Delta_r$ and $\Delta_y$. Requiring these singularities to be removable yields additional constraints on these functions~\cite{Ernst1976,Astorino:2022prj} which would need to be further explored.
	
	The Kerr metric belongs to this family. To see this, one can perform the following change of coordinates\footnote{We can see that the change of variables~\eqref{relcoord} is singular in the Schwarzschild limit $a \to 0$. Nevertheless, one does indeed recover the Schwarzschild black hole by performing the change of variables~\eqref{relcoord}, the choice~\eqref{KerrFunc} and \emph{then} taking the $a \to 0$ limit. One can note that such parameter singularities are common when one recovers common solutions from the Plebanski-Demianski family of spacetimes~\cite{Plebanski:1976gy}.}
	\begin{equation}
	\label{relcoord}
		y = a \cos{\theta} \,, \qquad \varphi = \frac{\phi}{a} \,, \qquad \tau = t - a \phi\,,
	\end{equation}
	which relates the coordinates $(\tau, r, y, \varphi)$ for the Boyer-Lindquist coordinates $(t, r, \theta, \phi)$. Then, the Kerr metric is recovered by setting
	%
	\begin{equation}
		\Delta^{\text{Kerr}}_r =  r^2 + a^2 - 2 Mr \,, \qquad \Delta^{\text{Kerr}}_y = a^2 - y^2 \,.
		\label{KerrFunc}
	\end{equation}
	%
	%
	%
Within this family, one can also find all the other representants of the Plebianski-Demianski family which are labeled by the NUT charge, the acceleration and the cosmological constant \cite{Carter:1968ks}. We shall consider these extended family in the last section.
The key interest in working with this family of geometries is that it is the most general set of geometries which possess the same hidden symmetries as the Kerr metric. These hidden symmetries are encoded in the so called Killing tower structure.

	To understand the structure, it is crucial to understand that all the symmetries of Kerr descend (can be canonically derived) from the existence of a principal tensor, i.e. a closed non-degenerate conformal Killing-Yano 2-form which is defined by the following equations:
	\begin{equation}
\nabla_{\mu} p_{\nu\alpha}  =  2 g_{\mu[\nu} h_{\alpha]} \,, \qquad \nabla_{[\mu} p_{\nu\alpha]}  =0 \,, \qquad h_{\alpha} = \frac{1}{3} \nabla_{\mu} p^{\mu}{}_{\alpha} \,.
\end{equation}
For all the members of the off-shell Kerr family, this tensor takes the form\footnote{Computations performed in this section have been checked by means of a SageMath notebook posted at\\ 
\url{https://nbviewer.org/url/relativite.obspm.fr/notebooks/Kerr_off_shell.ipynb}.}
	\begin{equation}
		p_{\mu\nu} \dd x^{\mu}\wedge \dd x^{\nu} = y \dd y \wedge (\dd \tau - r^2 \dd \varphi) - r \dd r \wedge (\dd \tau + y^2 \dd \varphi) \,.
	\end{equation}
	Using this object, one can construct the associated (dual) KY tensor, i.e. $f_{\mu\nu}= \frac{1}{2} \epsilon_{\mu\nu\alpha\beta} p^{\alpha\beta}$, which satisfies $\nabla_{(\mu} f_{\nu)\alpha} =0$ and which reads
	\begin{equation}
		f_{\mu\nu} \dd x^{\mu}\wedge \dd x^{\nu} = r \dd y \wedge (\dd \tau - r^2 \dd \varphi) + y \dd r \wedge (\dd \tau + y^2 \dd \varphi) \,.
	\end{equation}
	These two anti-symmetric tensors can then be combined to construct a conformal Killing tensor and a Killing tensor. The Killing tensor which will generate the Carter-like constant is given by
	\begin{align}
		K_{\mu\nu} \dd x^{\mu} \dd x^{\nu} & = f_{\mu\alpha} f_{\nu}{}^{\alpha} \dd x^{\mu} \dd x^{\nu} \nn \,, \\
		& = \frac{1}{\Sigma} \left[ y^2 \Delta_r (\dd \tau + y^2 \dd \varphi)^2 + r^2 \Delta_y (\dd \tau - r^2 \dd \varphi)^2 \right] + \Sigma \left[ \frac{r^2 \dd y^2}{\Delta_y} - \frac{y^2 \dd r^2}{\Delta_r} \right]  \,,
	\end{align}
	and satisfies $\nabla_{(\mu} K_{\nu\alpha)} =0$.
	%
	Finally, one can identify the killing vectors (KV) generating the time and azimuthal isometries from the above objects. Indeed, the so-called primary vector and the second vector can be derived from the principal tensor $p_{\mu\nu}$ and the Killing tensor $K_{\mu\nu}$ via the relations
	\begin{align}
		\xi^{\alpha} \partial_{\alpha }& = \frac{1}{3} \nabla_{\mu} p^{\mu\alpha}  \partial_{\alpha }= \partial_{\tau} \,, \\
		\chi^{\alpha}  \partial_{\alpha } & = - K^{\alpha}{}_{\mu} \xi^{\mu}  \partial_{\alpha } = \partial_{\varphi} \,.
	\end{align}
One recovers in this way the two commuting KVs of this stationary and axisymmetric spacetime. This elegant structure was initially identified and explored in \cite{Krtous:2006qy, Kubiznak:2006kt, Krtous:2008tb} when constructing higher dimensional rotating black hole geometries preserving the Kerr symmetries (see the review \cite{Frolov:2017kze}, page 44, section 3.2, for more details). This KOS family of four-dimensional geometries is uniquely defined by the existence of this Killing tower \cite{Krtous:2006qy}. Therefore, it naturally provides a new parametrization for Kerr-like objects which is selected solely on symmetry arguments. Let us now discuss the main novelty it provides compared to the parametrizations proposed so far.

	\subsection{Comparison with previous parametrizations}
	
	The search for a suitable parametrization of Kerr-like objects was initiated by Johannssen and Psaltis in \cite{Johannsen:2010xs}. Modifications to the initial JP proposal were studied in \cite{Cardoso:2014rha, Papadopoulos:2018nvd, Konoplya:2018arm, Carson:2020dez, Carson:2020iik, Yagi:2023eap, Salehi:2023eqy}, providing new parametrizations preserving some of the Kerr symmetries, in particular allowing for a Carter-like constant. In the following, we compare the KOS parametrization with the latest parametrizations proposed so far. 
	
	A general parametrization of Kerr-like objects was explored in \cite{Yagi:2023eap}. Consider the metric
	\begin{align}
	\rd s^2 = g_{tt} \rd t^2 + 2 g_{t\phi} \rd \phi + g_{rr} \rd r^2 + g_{\theta\theta} \rd \theta^2 + g_{\phi\phi} \rd \phi^2 \,,
	\end{align}
	where the metric functions read explicitly
	\begin{subequations}
	\begin{align}
		g_{tt} & = - \frac{\tilde{\Sigma} (A_5 - a^2 A^2_2 \sin^2{\theta})}{\rho^4}  \,, &g_{t\phi} & = \frac{a A_5 (A_5 - A_0) \tilde{\Sigma} \sin^2{\theta} }{\rho^4}\,, &g_{\theta\theta} & = \tilde{\Sigma}  \,,\\
		g_{\phi\phi} & = \frac{\tilde{\Sigma} \sin^2{\theta} A_5 (A^2_1 - a^2 A_5 \sin^2{\theta})}{\rho^4} \,, &g_{rr} & = \frac{\tilde{\Sigma}}{A_5}  \,,
	\end{align}
	\end{subequations}
	with
	\begin{align}
		\tilde{\Sigma} & = r^2 + a^2 \cos^2{\theta} + f(r) + g(\theta) \,, \\
		\rho^4 & = a^4 A^2_2 A_5 \sin^4{\theta} + a^2 \sin^2{\theta} (A^2_0 - 2 A_0 A_5 - A^2_1 A^2_2) + A^2_1 A_5 \,.
	\end{align}
The functions $A_i(r)$, $f(r)$ and $g(\theta)$ are arbitrary\footnote{The initial JP parametrization is recovered for $A_0 = A_1 A_2$.}. The Petrov type of this general geometry was studied in detail in \cite{Yagi:2023eap}, showing that, without further conditions, such geometry is not algebraically special, i.e. is of Petrov type I. Within this general family, the authors identified two restricted classes which admit a Carter-like constant. 
	\begin{itemize}
	\item The first class (hereafter Class~I) is given by imposing
	\begin{align}
	    A_0 = A_1 = r^2+a^2 \;, \qquad A_2 =1\;, \qquad f =g =0
	\end{align}
	leading to 
	\begin{subequations}
	\begin{align}
		g_{tt} & = - \frac{ (A_5(r) - a^2 \sin^2{\theta})}{\Sigma} \,, &g_{t\phi}  &= \frac{a ( r^2 + a^2 - A_5(r)) \sin^2{\theta} }{\Sigma} \,, &g_{\theta\theta}  &= \Sigma \,, \\
        g_{\phi\phi} & = \frac{(r^2+a^2)^2 - a^2 A_5(r) \sin^2{\theta})}{\Sigma} \,, &g_{rr}  &= \frac{\Sigma}{A_5(r)} \,,
	\end{align}
	\end{subequations}
	where $\Sigma = r^2 + a^2 \cos^2{\theta} $.
	A direct comparison shows that this class is a sub-case of the KOS family such that
	\be
	\label{Kerrr}
	\Delta_r(r) = A_5(r) \,,\qquad \Delta_y(y) = \Delta^{\text{Kerr}}_y(y) = a^2 - y^2 = a^2 \sin^2{\theta} \,.
	\ee
	This family of geometries only contains radial deformations while remaining blind to any polar corrections. Nevertheless, it contains a large class of corrections studied so far, in particular all the regular black holes built by deforming the function $\Delta^{\text{Kerr}}_r(r) = r^2 + a^2 - 2 M r $ to $\Delta_r(r) = r^2 + a^2 - 2 M(r) r $ with a suitable function of $M(r)$. As shown in \cite{Yagi:2023eap}, all members of this class are of Petrov type D. The fact that all these geometries belong to the KOS family implies that each of them admits not only a Killing tensor, but a killing tensor which descends from a Killing-Yano tensor. This is coherent with the known theorems relating the existence of KY two-form and the Petrov type of a given geometry. Indeed, only Petrov type D, N, and O geometries are equipped with a KY two form \cite{Batista:2020cto}. 
	\item The second restricted Class II corresponds to
	\be
	A_0 = A_1 = r^2+a^2 + f(r), \qquad A_2 =1,\qquad g =0
	\ee
	while all the functions $A_5(r)$ and $f(r)$ are left free. As shown in \cite{Yagi:2023eap}, these geometries are of Petrov type I. As an example, one can obtain the Kerr-Sen black hole by setting all the free functions to zero while imposing $f(r) = \alpha r$ and $A_5(r) = r^2 + a^2 - 2 M r + \alpha r$. \textit{Since this class of geometries allows for corrections to the function $\Sigma= r^2 + a^2 \cos^2{\theta}$, and is of Petrov type I, it cannot be covered by the KOS family which is by construction Petrov type D.} Therefore, representants of this second class provide interesting examples of non-algebraically special geometry which admits a Killing tensor but no Killing-Yano twoform.  
	\end{itemize}
	Therefore, compared to the parametrization studied in \cite{Yagi:2023eap}, the KOS family allows us to single out the Petrov type D geometries belonging to the restricted class I, but also a whole new set of geometries admitting polar deformations while still equipped with a Killing-Yano tensor. 
	
	To conclude, we point that a parametrization of Kerr-like objects with general polar and radial deformations have been studied in \cite{Konoplya:2016jvv}. However, the allowed freedom is so general that the considered geometries fail to be algebraically special and thus preserve the Kerr symmetries tied to the Petrov type D. The polar deformation $\Delta_y(y)$ of the KOS family can be understood as the only allowed polar deformation selected by the Killing tower structure. We are now ready to study the geodesic motion and the photon ring structure for this KOS family.


	\section{Geodesic motion and photon ring}
	
	\label{geod}
	
	In this section, we study the geodesic motion on a general KOS geometry in the Hamiltonian formalism. We derive the Carter-like constant and discuss how the deformations $(\Delta_r(r), \Delta_y(y))$ affect the radial and polar motion of test particles. 
	Then, we study the unstable radial and polar positions for the geodesic motion and provide close formula for the allowed range of the radial and polar coordinates for such unstable equilibriums.

	\subsection{Generalized Carter constant and geodesic equations}
	
	Consider the general KOS metric (\ref{offshell}). The phase space of the geodesic motion on this background is described by the coordinates $\tau$, $r$, $y$, $\varphi$ and the impulsions $p_\tau$, $p_r$ $p_y$ and $p_\varphi$.
	%
	%
The canonical commutation relations read
	\begin{align}
		\poissonbracket{\tau}{p_\tau} &= 1 \,, & \poissonbracket{r}{p_r} &= 1 \,, & \poissonbracket{y}{p_y} &= 1 \,, & \poissonbracket{\varphi}{p_\varphi} &= 1 \,,
	\end{align}
	while all other Poisson brackets are zero. 
	Notice that
	\be
	\label{mom}
	p_r = \frac{\Sigma}{\Delta_y} p^y \qq{and} p_r = \frac{\Sigma}{\Delta_r} p^r \,.
	\ee
	The geodesic Hamiltonian generating the dynamics of test particles is given by
	\begin{equation}
	\label{ham}
		H = \frac{1}{2}g^{\mu\nu} p_{\mu} p_{\nu} = \frac{1}{2\Sigma} \left[ - \frac{1}{\Delta_r} \left( r^2 p_\tau + p_\varphi \right)^2 + \frac{1}{\Delta_y} \left( y^2 p_\tau -  p_\varphi \right)^2 + \Delta_r p_r^2 + \Delta_y p_y^2 \right] \,.
	\end{equation}
	As expected for a stationary and axi-symmetric metric, the variables $(\tau, \phi)$ are cyclic, such that $p_{\tau}$ and $p_\varphi$ are constants of motion, i.e. $\poissonbracket{p_\tau}{H} = 0$ and $\poissonbracket{p_\varphi}{H} = 0$. They correspond to the conserved charges induced by the time and azmutal Killing vectors $\xi^{\mu} \partial_{\mu} = \partial_{\tau}$ and $\chi^{\mu} \partial_{\mu} = \partial_{\varphi}$ given by
	\be
	E = - \xi^{\mu} p_{\mu} = -p_{\tau } \,\qquad L = \chi^{\mu} p_{\mu} = p_{\varphi} \,.
	\ee 
Notice that since our coordinates $(\tau, \varphi)$ are not the standard BL coordinates $(t,\phi)$, the above charges $(E, L)$ do not coincide with the standard energy and angular momentum derived in the BL coordinates\footnote{Using the change of coordinates (\ref{relcoord}), one finds that the two Killing vectors are given by $\xi^{\mu} \partial_{\mu} = \partial_{t}$ and $\chi^{\mu} \partial_{\mu} = a(\partial_{\phi} + a \partial_t)$ in terms of BL coordinates. Introducing the momenta $(p_t, p_{\phi})$ and the energy $E^{\text{BL}} = - p_t$ and angular momentum $L^{\text{BL}} = p_{\phi}$ associated to the BL coordinates, one obtains therefore
\be
\label{rrel}
E = E^{\text{BL}} \qquad L = a \left( L^{\text{BL}} - a E^{\text{BL}}\right)
\ee
which will be useful when comparing to the Kerr case.}.
Then, the existence of Killing tensor (descending from the Killing-Yano twoform) provides another conserved charge quadratic in the momenta. This generalized Carter constant is given by
	\begin{align}
	\label{Carter}
		K = K^{\mu\nu}p_{\mu}p_{\nu} = \frac{1}{\Sigma} \left[\frac{y^2}{\Delta_r} \left( r^2 p_\tau + p_\varphi \right)^2 + \frac{r^2}{\Delta_y} \left( y^2 p_\tau -  p_\varphi \right)^2 - y^2 \Delta_r p_r^2 + r^2 \Delta_y p_y^2\right] \,.
	\end{align}
		One can verify that $\poissonbracket{K}{H} = 0$, which means that the quantity $K$ is indeed conserved along a geodesic. 
It is useful for the following to simplify this constant of motion. To that end, we use the trivial constant of motion given by the mass $m$ of the particle, i.e. its hamiltonian, such that $2H = p_\mu p^\mu = - m^2$. Using (\ref{mom}) and (\ref{ham}), one can simplify as:
	\begin{equation}
		K = \frac{1}{\Delta_y} \left[ \left( y^2 p_\tau -  p_\varphi \right)^2 + \Sigma^2 (p^y)^2\right]  +  y^2 m^2  = \frac{1}{\Delta_y} \left[ \left( y^2 E +L \right)^2 + \Sigma^2 (p^y)^2\right]  +  y^2 m^2
		\,,
	\end{equation}
or equivalently as
			\begin{equation}
		K  = \frac{1}{\Delta_r} \left[ (r^2p_{\tau} + p_{\varphi})^2 - \Sigma^2 (p^r)^2\right] + r^2 m^2 = \frac{1}{\Delta_r} \left[ (r^2 E -L)^2 - \Sigma^2 (p^r)^2\right] + r^2 m^2
		\,.
	\end{equation}
	%
	%
	%
	These four constants of motion $(p_{\tau}, p_{\phi}, K, H)$ commute with each other, which imply that only 4 out of the initial 8 degrees of freedom are left in the phase space, thus making the geodesic motion integrable. 
	With the phase space at hand, we can now derive the dynamics of a test particle. 
	
	Let us introduce an affine parameter denoted $\lambda$. 
	Hamilton's equations for $\tau$, $r$, $y$ and $\varphi$ are given by
	\begin{subequations}
		\begin{align}
		\Sigma \dv{\tau}{\lambda} &= - \frac{r^2 (r^2 p_\tau + p_\varphi)}{\Delta_r} + \frac{y^2 (y^2 p_\tau - p_\varphi)}{\Delta_y} \,,\\
			\label{geoo}
		\Sigma \dv{r}{\lambda} & = \Delta_r p_r \,, \\
		\Sigma \dv{y}{\lambda} & = \Delta_y p_y \,, \\
		\Sigma \dv{\varphi}{\lambda} &= - \frac{r^2 p_\tau + p_\varphi}{\Delta_r} - \frac{y^2 p_\tau - p_\varphi}{\Delta_y}  \,,
	\end{align}\label{eq:propag-phi}
	\end{subequations}
These equations can be further simplified be expressing them in terms of the conserved charges $(E, L, K, m)$. One can extract the expression for $p_r$ and $p_y$ from the expressions for $m^2$ and $K$, yielding	
    \begin{subequations}
    \begin{align}
		\Sigma \dv{\tau}{\lambda} &= \frac{r^2 (r^2 E - L)}{\Delta_r} - \frac{y^2 (y^2 E + L)}{\Delta_y}\,,\\
			\label{geopot}
		\Sigma \dv{r}{\lambda} & = \pm \sqrt{V_r(r)} \,, \\
		 \Sigma \dv{y}{\lambda} & = \pm \sqrt{V_y(y)} \,, \\
		\Sigma \dv{\varphi}{\lambda} & =  \frac{r^2 E - L}{\Delta_r} + \frac{y^2 E + L}{\Delta_y} \,,
	\end{align}\label{eq:propag-phi}
    \end{subequations}
where the two potentials $V_r(r)$ and $V_y(y)$ are given by
    \begin{subequations}
    \begin{align}
	\label{potgeo1}
		V_r(r) &= (r^2 E - L)^2 - \Delta_r (K + r^2 m^2) \,,\\
			\label{potgeo2}
		V_y(y) &= \Delta_y (K - m^2 y^2)   -(y^2 E + L)^2 \,.
	\end{align}
    \end{subequations}
	This provides the general set of first order equations describing the geodesic motion on an arbitrary representant of the KOS family. 
	The two potentials (\ref{potgeo1}) and (\ref{potgeo2}) govern respectively the radial and polar motion of the test particle\footnote{In order to compare (\ref{potgeo1}) and (\ref{potgeo2}) to the standard potentials involved in the radial and polar motion around Kerr potentials, it is useful to write the equation for $\theta(\lambda)$ instead of $y(\lambda)$, using that $\theta =  \text{arcos}{(y/a)}$ such that
	\begin{equation*}
	\left( \frac{\rd \theta}{\rd \lambda}\right)^2 =  \frac{1}{a^2 -y^2} \left( \frac{\rd y}{\rd \lambda}\right)^2 \,.
	\end{equation*}
This leads to the following equation:
	\begin{equation*}
	\Sigma^2 \left( \frac{\rd \theta}{\rd \lambda}\right)^2 = \Theta (y) \,, \qquad \Theta (y) = \frac{\Delta_y (K - m^2 y^2)   -(y^2 E + L)^2}{a^2 - y^2} \,.
	\end{equation*}
One recovers the Kerr potentials using (\ref{rrel}) and (\ref{Kerrr}), i.e. 
	\begin{align*}
	V_r (r) & = \cR (r) = \left[ (r^2+a^2)E^{\text{BL}} - a L^{\text{BL}}\right]^2 - \Delta (r^2 m^2 + K) \,,\\
	V_y(y) &= \Theta(\theta) = K - \left( \frac{L^{\text{BL}}}{\sin{\theta}} - a E^{\text{BL}} \sin{\theta}\right)^2 - m^2 a^2 \cos^2{\theta} \,.
	\end{align*}}. The classification of the geodesic motion on the Kerr black hole has a long story \cite{Wilkins, Schmidt:2002qk, Drasco:2003ky, Bardeen, Fujita:2009bp, OShaughnessy:2002tbu, Mino:2003yg, Dexter},  a full characterization of the set of allowed trajectories being still an active research field \cite{Gralla:2019ceu, Compere:2021bkk}. See \cite{Druart:2023mba} for a detailed discussion. The different Kerr geodesics can be classified in terms of the range of the constants of motion $(E, L, K, m)$ and the spin $a$\footnote{Notice that it is standard to introduce another constant $Q$ which is well suited to study the equatorial plane, i.e. $y=0$ or $\theta=\pi/2$. In our case, this alternative Carter constant can be written
	\begin{align}
	\label{CC}
	Q = \frac{V_y(0)}{a^2} = \frac{K \Delta_y(0) - L^2}{a^2} = \frac{K \Delta_y(0) - a^2 (L^{\text{BL} }- a E^{\text{BL}})^2}{a^2}
	\end{align}
	which for Kerr reduces indeed to
	\be
	Q = K - (L^{\text{BL} }- a E^{\text{BL}})^2
	\ee}. A first condition for causal geodesics is that 
	\be
	V_r(r) \geqslant 0 \qquad \text{and} \qquad V_y(y) \geqslant 0
	\ee
at any point $(r,y)$ of the geodesic flow. Assuming that $\Delta_y(y) \geqslant 0$, one sees that the positivity of $V_y(y)$ implies that the generalized Carter constant is always positive or vanishing, i.e.
\be
K \geqslant 0 \,.
\ee
In the Kerr case, where $\Delta^{\text{Kerr}}_y(y) = a^2 -y^2 = a^2 \sin^2{\theta} \geqslant 0$, this condition is satisfied and one recovers the positivity of the Carter constant. For the Kerr black hole, one can split the study between the trajectories with $L=0$ and the ones with $L\neq 0$. While the first are allowed to reach the rotation axis and cross it, the later family never reach it and are either confined between two critical polar angles or constrained to stay at a given angle $\theta$. One can then distinguish several families of trajectories with specific qualitative behavior depending on the range of the constants $(E, L, K, a)$. See \cite{Chand83,ONeil95, Gralla:2019ceu, Compere:2021bkk} for details.

Obviously, such classification cannot be done in our case without selecting concrete functions $\Delta_r(r)$ and $\Delta_y(y)$. Nevertheless, one can derive the general expressions for the equilibrium positions of the radial and polar motion for a general KOS geometry and study the set of critical spherical photon trajectories which characterize the photon ring.

	\subsection{Photon ring of the Kerr off-shell metrics}
	
	The photon ring is generated by photon trajectories
	that asymptote (unstable) bound photon orbits --- the so-called \enquote{photon spherical orbits} to be discussed below.
	Thus, let us fix the particle mass to zero, i.e. $m = 0$ and let us focus on the null geodesics with $E \neq 0$. This allows one to introduce the reduced constants of motion given by
	\begin{equation}
		\ell = \frac{L}{E} \qq{and} k = \frac{K}{E^2} \,.
	\end{equation}
Imposing $m=0$, and using the new constants, the reduced geodesic equations read\footnote{Since we are focusing on photons that are reaching a distant observer, we may set $|E| = E$ in the following equations, given that one has necessarily $E>0$ for such photons (only photons confined to the ergoregion can have $E\leq 0$).}
    \begin{subequations}
    \begin{align}
		\frac{\Sigma}{E} \dv{\tau}{\lambda} &= \frac{r^2 (r^2  - \ell)}{\Delta_r} - \frac{y^2 (y^2  + \ell)}{\Delta_y}\,,
		\\
		\label{geopot1}
		\frac{\Sigma}{E} \dv{r}{\lambda} & = \pm \sqrt{\cV_r(r)} \,, \\
		 \label{geopot2}
		 \frac{\Sigma}{E}  \dv{y}{\lambda} & = \pm \sqrt{\cV_y(y)} \,, \\
		 \label{geopot3}
		\frac{\Sigma}{E}  \dv{\varphi}{\lambda} & =  \frac{r^2  - \ell}{\Delta_r} + \frac{y^2  + \ell}{\Delta_y} \,,
	\end{align}
    \end{subequations}
where the two potentials $\cV_r(r)$ and $\cV_y(y)$ are given by
    \begin{subequations}
    \begin{align}
	\label{potgeo11}
		\cV_r(r) &= \frac{V_r(r)}{E} =  (r^2  - \ell)^2 - \Delta_r k \,,\\
			\label{potgeo22}
		\cV_y(y) &= \frac{V_y(y)}{E} = \Delta_y k  -(y^2  + \ell)^2 \,.
	\end{align}
    \end{subequations}
In the following, we identify the spherical photon orbits which play the role of the critical orbits beyond which photon can escape to infinity. The photon orbits which are infinitely close to the critical orbits are the ones generating the image of the photon ring. They are characterized by a set of parameters $(\ell_c, k_c)$ which we now determine.

		\subsubsection{Radial motion and spherical photon orbits}
		
		\label{crit}
		
		A \textit{photon spherical orbit}\footnote{Here, the denomination \enquote{spherical} should be understood in the topological sense: indeed, the metric induced by $(\tau = \text{cst}, r = \text{cst})$ on~\eqref{offshell} is not the metric of a 2D sphere in general.} is a null geodesic at fixed value, say $r_0$, of the coordinate $r$
		\cite{Teo:2003ltt}. 
		
		Spherical photon orbits obey $\dv*{r}{\lambda}=0$ for all values of $\lambda$. According to~\eqref{geopot1}, this implies
		\be
		\cV_r(r_0) =0 \,,\qquad  \cV'_r(r_0) =0 \,.
		\ee
		These conditions translate into two equations for three unknowns $(r_0, \ell_c, k_c)$ given by
		\begin{subequations}
		\begin{align}
		(r^2_0 - \ell_c)^2 - \Delta_r(r_0) k_c =0 \,, \\
		4 r_0 (r^2_0 - \ell_c) - \Delta'_r(r_0) k_c =0 \,.
		\end{align}
		\end{subequations}
		Given some $r_0$, we look for the solutions $(\ell_c, k_c)$ labeling to allowed photon trajectories. Let us assume that $\Delta'_r(r_0) \neq 0$. Then, one can solve for $k$:
		\begin{align}
		k_c  = \frac{4 r_0 (r^2_0 -\ell_c)}{\Delta'_r(r_0)} \,.
		\end{align}
		Injecting this solution in the last equation, one obtains 
			\begin{align}
		 (\ell_c - r^2_0) \left( \ell_c - r^2_0 + \frac{4r_0 \Delta_r(r_0)}{\Delta'_r(r_0)}\right) & =0 \,,
		\end{align}
		which has two solutions. The first one is given by $(\ell_c, k_c)  = (r^2_0, 0)$. Now injecting $k=0$ in the polar potential, one finds $V_y(y) = - (y^2+ \ell)^2$ which has to be positive for all $y$. Thus, for $y=0$, one has $\ell_c =0$ which implies $r_0 =0$ and one finds that this solution is not admissible because it is clearly under the outermost horizon. The second solution $(\ell_c,k_c)$ corresponds to 
		\be
		\label{sol}
		\boxed{\ell_c = r_0 \left( r_0 -  \frac{4 \Delta_r(r_0)}{\Delta'_r(r_0)} \right) \,,\qquad k_c = \frac{16 r^2_0 \Delta_r(r_0)}{\Delta'_r(r_0)^2}} \,.
		\ee
		Notice that as expected, $(\ell_c, k_c)$ do not depend on the free function $\Delta_y$.
		Now, in order to fully characterize the locus of points allowing for spherical photon orbits, we have to further impose that $\mathcal{V}_y(y) \geqslant 0$. 
		
		At the equator, i.e. at $y=0$, one has
		\be
		\mathcal{V}_y(0) \geqslant 0 \qquad  \rightarrow \qquad \Delta_y(0)k_c  \geqslant \ell^2_c \,.
		\ee
The above condition
		 translates into the following inequality
		\begin{align}
		\label{ph}
	\frac{r^2_0}{\Delta'_r(r_0)^2} \left[ 16 \Delta_r(r_0) \Delta_y(0) - (r_0 \Delta_r'(r_0) - 4 \Delta_r(r_0))^2 \right] \leqslant 0
		\end{align}
		At this stage, one needs to pick up a given function $\Delta_r(r)$ to find the associated positions $r_0$ satisfying this inequality. In the case of Kerr, one finds that the condition reduces to a cubic polynomial equation from which one can extract the range of $r_0$. In the following, we shall assume that the function   $\Delta_r(r)$ is such that the largest set of values of $r_0$ allowed by this inequality belong to an interval 
		\begin{align}
		r_0 \in \interval{r^{-}_{\text{ph}}}{r^{+}_{\text{ph}}} \,,
		\end{align}
		where we have 
		\begin{align}
		r^{-}_{\text{ph}} < r^{+}_{\text{ph}}\;, \qquad  \Delta_r (r^{-}_{\text{ph}}) >0 \qquad \text{and} \qquad  \Delta_r (r^{+}_{\text{ph}}) >0 \,.
		\end{align}
		This concludes the study of the spherical photon orbits in the KOS geometry. Given one representant of the family, one can determine the couple $(\ell_c, k_c)$ labeling these critical orbits and the radial positions $r^{\pm}_{\text{ph}}$ for which spherical orbits exist. At this stage, we are left with studying the polar motion.

	
		\subsubsection{Polar motion and stable equilibrium positions}
		
		\label{polarmo}
	
In order to capture the different allowed polar motions, it is instructive to recast the polar equation of motion into an equation for a 1d particle in a potential. By setting $\dd{\lambda'} = \abs{E}\sqrt{\Delta_y}\dd{\lambda}/\Sigma$, using the fact that $\Delta_y \geq 0$, this equation can be recast into
	\begin{equation}
		\Big(\dv{y}{\lambda'}\Big)^2 + \Theta(y) = k  \qquad \text{with} \qquad \Theta(y) = \frac{(y^2 + \ell)^2}{\Delta_y} \,.
		\label{eq:pot-eq-y}
	\end{equation}
	This can be interpreted as a one-dimensional motion in a potential $\Theta(y)$ and total energy $k$. The allowed trajectories therefore have to satisfy
	\be
	0 \leqslant\Theta(y) \leqslant k \,.
	\ee
	The derivatives of the potential are given by
	\begin{subequations}
	\begin{align}
		\Theta'(y) &= \frac{y^2+\ell}{\Delta^2_y} \left(  4 y\Delta_y  - (y^2+\ell) \Delta'_y \right) \\
		\Theta''(y) &= \frac{8y^2}{\Delta_y} + \frac{4(y^2+ \ell)}{\Delta_y^2} (\Delta_y - 2 y \Delta_y') + \frac{(y^2+ \ell)^2}{\Delta_y^3} (2\Delta_y'^2 - \Delta_y \Delta_y'')
	\end{align}	\label{eq:extrema-Vy}
	\end{subequations}
	Let us denote the different possible roots of $\Delta_y(y)$ by $y_{\ast}^\pm$. These positions bound the available domain for $y$. In the case of Kerr, they correspond to the axis of rotation.
	
	In order to understand the polar motion we need to identify the equilibriums. A polar equilibrium at $y=y_e$ satisfies $\Theta(y_e)=0$ and $\Theta' (y_e) = 0$. Depending on the stability of the various equilibriums, several cases arise. For simplicity, we assume that $\Delta_y$ is concave at $y=0$: this implies $\Delta_y''(0)<0$.
	\begin{itemize}
	    \item First, we see that $y_e = 0$ is always a solution: indeed, since $\Delta_y$ is even, one has $\Delta_y'(0)=0$. One then obtains 
	    \begin{equation}
	        \Theta_y''(0) = \frac{\ell}{\Delta_y(0)} \left(4 - \ell \frac{ \Delta_y''(0)}{\Delta_y(0)}\right) \,.
	    \end{equation}
	    The position is stable only if $\Theta_y''(0) > 0$. This corresponds to
	    \begin{equation}
	        \ell > 0 \qq{or} \ell < \frac{4\Delta_y(0)}{\Delta_y''(0)} \,.
	    \end{equation}
	    If $\ell=0$, stability depends on the third derivative of $\Theta$.
	    \item Second, we see that when $\ell < 0$, $y_e = \pm\sqrt{-\ell}$ are two solutions as well. When they exist, these solutions are always stable, since $\Theta''(\pm\sqrt{-\ell}) > 0$.
	\end{itemize}
	Depending on the exact expression of $\Delta_y$, there might be other positions for which $\Theta'=0$. For simplicity, we assume that this is not the case. The possible movements then depend on the value of $\ell$:
	\begin{itemize}
	    \item if $\ell>0$, there are two positions $\pm y_m$ such that $\Theta(\pm y_m) = k$ and the polar motion corresponds to oscillations between $\pm y_m$;
	    \item if $\flatfrac{4\Delta_y(0)}{\Delta_y''(0)} < \ell< 0$, the movement depends on the value of $k$: either the equation $\Theta(y)=k$ has two solutions $\pm y_m$ and the motion corresponds to oscillations around the equatorial plane between these two values, or it has four solutions  $\pm y_m$ and $\pm y_v$ (with $0<y_v<y_m$) and we can have oscillations either in $\interval{-y_m}{-y_v}$ and $\interval{y_v}{y_m}$;
	    \item if $\ell < \flatfrac{4\Delta_y(0)}{\Delta_y''(0)}$, the movement depends on the relative values of $y_\ast^\pm$ and the roots of $\Theta(\pm y_m) = k$ --- one should note that this case does not exist in the Kerr limit because $y_\ast^\pm = \pm a$ in this case;
	    \item if $\ell =0$, the motion depends on the exact formulation of $\Delta_y$ through the third derivative of $\Theta$.
	\end{itemize}

Let us recall that for the critical orbits relevant for the photon ring image, the parameters $(\ell_c, k_c)$ are already fixed in term of the radial position $r_0$ by the solution (\ref{sol}) for the spherical photon orbits. 
Having discussed the radial and polar motion of spherical photon orbits in the KOS family, we now turn to the black hole image and the derivation of the critical curve.
			
	\subsection{Screen angular coordinates and the critical curve}
		\label{sec:shadow}
		
	We now derive the projected image of the photon ring as seen on the screen of an asymptotic observer.  To that end, we consider an observer at fixed spatial coordinates given by $(r_\mathcal{O}, y_\mathcal{O}, \varphi_\mathcal{O})$. We assume that $r_\mathcal{O} \gg r_\mathrm{h}$ while $y_{\cO}$ encodes the inclination of the observer w.r.t the black rotation axis, i..e $y_{\cO}=0$ corresponds to an asymptotic observer in the equatorial plane of the black hole.  
	
	We further assume that the observer has zero angular momentum. The orthogonal frame associated to this observer is given by
	\begin{subequations}\label{eq:ZAMO-frame}
	\begin{align}
		\bm{e}_{(\tau)} &= \sqrt{\frac{r^4 \Delta_y - y^4 \Delta_r}{\Sigma \Delta_r \Delta_y}} \Big( \bm{\partial}_\tau + \frac{y^2 \Delta_r + r^2 \Delta_y}{r^4 \Delta_y - y^4 \Delta_r} \bm{\partial}_\varphi \Big) \,,\\
		\bm{e}_{(r)} &= \sqrt{\frac{\Delta_r}{\Sigma}} \bm{\partial}_r \,,\\
		\bm{e}_{(y)} &= \sqrt{\frac{\Delta_y}{\Sigma}} \bm{\partial}_y \,,\\
		\bm{e}_{(\varphi)} &= \sqrt{\frac{\Sigma}{r^4 \Delta_y - y^4 \Delta_r}} \bm{\partial}_\varphi \,.
	\end{align}
	\end{subequations}
	which satisfies the properties that the observer's wordline is normal to the hypersurfaces of constant time $\tau$ and whose 4-velocity is orthogonal to $\bm{\partial}_\varphi$.
	Now, we assume that this observer is equipped with a screen centered on the direction to the black hole such that $\bm{e}_{(r)}$ is normal to the screen. In the large $r$ limit, this frame reduces to
		\begin{align}
		\bm{e}_{(\tau)} &= \bm{\partial}_\tau \,, & \bm{e}_{(r)} &= \bm{\partial}_r \,, & \bm{e}_{(y)} &= \frac{\sqrt{\Delta_y(y_{\cO})}}{r_\mathcal{O}} \bm{\partial}_y \,, & \bm{e}_{(\varphi)} &= \frac{1}{r_\mathcal{O}\sqrt{\Delta_y(y_{\cO})}} \bm{\partial}_\varphi \,.
	\end{align}
	such that $\bm{e}_{(\tau)}$ is the unit vector orthogonal to constant-$\tau$ hypersurfaces.
	
	Now let us consider the photon trajectories generating the image of the black hole on the observer's screen. The observer will collect photons whose trajectories satisfy
	\begin{subequations}
	\begin{align}
	\cV_r(r_\mathcal{O}) &= (r^2_{\cO}  - \ell)^2 - \Delta_r(r_{\cO}) k \geqslant 0 \,,\label{eq:cond-Vr} \\
	 \cV_y(y_\mathcal{O}) & = \Delta_y(y_{\cO}) k - (y_{\cO}^2 + \ell)^2 \geqslant   0 \,.\label{eq:cond-Vy}
	\end{align}
	\end{subequations}
At large $r$, one has always $\mathcal{\cV}_r(r_{\cO}) \sim r_{\cO}^4$, which implies that~\eqref{eq:cond-Vr} is always satisfied. On the other hand,~\eqref{eq:cond-Vy} translates into a condition on the parameters $\ell$ and $k$ at a given observer's inclination $y_{\cO}$.
A photon that follows a geodesic $\Gamma$ has the following 4-momentum at the location of $\mathcal{O}$:
	\begin{equation}
		\bm{p} = p^\tau \bm{e}_{(\tau)} + p^r \bm{e}_{(r)} + \frac{p^y r_\mathcal{O}}{\sqrt{\Delta_y(y_{\cO})}} \bm{e}_{(y)} + p^\varphi r_\mathcal{O} \sqrt{\Delta_y(y_{\cO})} \bm{e}_{(\varphi)} = E (\bm{e}_{(\tau)} + \bm{n})\,,
	\end{equation}
	with $\bm{n}$ a unit spacelike vector that describes the photon's direction of propagation, seen from the point of view of $\mathcal{O}$, i.e. 
	\be
		\bm{n} =  p^r \bm{e}_{(r)} + \frac{p^y r_\mathcal{O}}{\sqrt{\Delta_y(y_{\cO})}} \bm{e}_{(y)} + p^\varphi r_\mathcal{O} \sqrt{\Delta_y(y_{\cO})} \bm{e}_{(\varphi)} \,.
	\ee
	As seen from the observer, the projected momenta is given by $-\bm{n}$. 
The relevant information for the photon ring image is carried  by the projected momenta of the photon on the observer's screen. Splitting the vector $-\bm{n} $ as $-\bm{n}  = - \bm{n}_{\perp} - \bm{n}_{\parallel}$, one has 
	\begin{equation}
	\label{proj}
 -\bm{n_\parallel} = - \frac{p^y r_\mathcal{O}}{E\sqrt{\Delta_y(y_{\cO})}} \bm{e}_{(y)} - \frac{p^\varphi r_\mathcal{O}\sqrt{\Delta_y(y_{\cO})}}{E}  \bm{e}_{(\varphi)} \,.
	\end{equation}
Now, let us introduce dimensionless Cartesian coordinates $(\alpha, \beta)$ on the screen such that the black hole's rotation axis appears as the $\beta$ axis. We then have
	\begin{equation}
		 -\bm{n_\parallel} = \alpha \bm{e}_{(\alpha)} + \beta \bm{e}_{(\beta)} \qq{with} \bm{e}_{(\alpha)} = \bm{e}_{(\varphi)} \qq{and} \bm{e}_{(\beta)} = - \bm{e}_{(y)} \,,
	\end{equation}
such that comparing to (\ref{proj}), one finds the following relation between the couple $(\alpha, \beta)$ and the components of the projected photon momenta:
	\begin{equation}
		\alpha = - \frac{p^\varphi r_\mathcal{O}\sqrt{\Delta_y(y_{\cO})}}{E}  \qq{and} \beta = \frac{p^y r_\mathcal{O}}{E\sqrt{\Delta_y(y_{\cO})}} \,.
	\end{equation}
	Now, using~\eqref{geopot3} and~\eqref{geopot2} to express $p^y$ and $p^\varphi$, one obtains
	\begin{equation}
		\alpha = -  \frac{r_\mathcal{O} \sqrt{\Delta_y(y_{\cO})}}{ E (r^2_{\cO} + y^2_{\cO})} \Big(\frac{r_\mathcal{O}^2 E - L}{\Delta_r(r_{\cO})} + \frac{y_\mathcal{O}^2 E + L}{\Delta_y(y_{\cO})}\Big) \,,\qquad 
		\beta = \pm \frac{r_{\cO}}{r^2_{\cO} + y^2_{\cO}} \left( k - \frac{(y_\mathcal{O}^2  + \ell)^2}{\Delta_y(y_{\cO})}\right)^{1/2}\,.
	\end{equation}
	Finally, taking the large $r_\mathcal{O}$ limit, the above relations reduce to
	\begin{subequations}\label{eq:coords-screen}
	\begin{align}
		\alpha &= - \frac{\sqrt{\Delta_y(y_{\cO})}}{r_\mathcal{O}} \Big(1 + \frac{y_\mathcal{O}^2 + \ell}{\Delta_y(y_{\cO})}\Big) \,,\\
		\beta &= \pm \frac{1}{r_\mathcal{O}} \left( k - \frac{(y_\mathcal{O}^2  + \ell)^2}{\Delta_y(y_{\cO})}\right)^{1/2} \,.
	\end{align}
	\end{subequations}
These equations relates the coordinates $(\alpha, \beta)$ of the impact of the photon on the screen with its parameters $(\ell, k)$ and the position and inclination of the observer $(r_{\cO}, y_{\cO})$. Now, the photon ring is generated by critical photons orbits which have parameters $(\ell,k)$ infinitely close to the critical parameters  $(\ell_c, k_c)$ characterizing spherical photon orbits\footnote{Nonspherical closed photon orbits around the black hole could exist and also contribute to the photon ring. A nonspherical photon orbit exists between a perihelion $r_\mathrm{p}$ and aphelion $r_\mathrm{a}$ if and only if
	\begin{equation*}
	    \mathcal{V}_r(r_\mathrm{p}) = \mathcal{V}_r(r_\mathrm{a}) = 0 \qq{and} \mathcal{V}_r(r) \geq 0 \qq{when} r \in \interval{r_\mathrm{p}}{r_\mathrm{a}} \,.
	    \label{eq:conds-nonspherical}
	\end{equation*}
	For the Kerr metric, this situation in investigated in~\cite{gourgoulhon}: due to the roots system of the potential, no such nonspherical orbit is found. For the Kerr off-shell family, despite careful investigation, we found no expression of $\Delta_r$ that allows for such orbits. This is the case in particular for all examples studied in this work.}. Therefore, these geodesics remain trapped in the region $r_0 \in \interval{r_\mathrm{ph}^+}{r_\mathrm{ph}^-}$ before escaping to infinity and eventually reaching the observer's screen.
	The black hole shadow is obtained as the image of these geodesics on a screen far away from the black hole. Therefore, using the expressions (\ref{sol}) for the critical parameters $(\ell_c, k_c)$ derived in Section~{\ref{crit}}, the cartesian coordinates $(\alpha_c, \beta_c)$ take the following form:
	\begin{subequations}
	\begin{align}
	\alpha_c & = - \frac{1}{r_{\cO}\sqrt{\Delta_y(y_{\cO})}} \left[   \Sigma(r_0, y_\mathcal{O}) + \Delta_y (y_{\cO}) - \frac{4 r_0\Delta_r(r_0)}{\Delta'_r(r_0)} \right]\,,\\
	\beta_c & = \pm  \frac{1}{r_{\cO}\sqrt{\Delta_y(y_{\cO})}} \left[ -16r_0^2 \frac{\Delta_r(r_0)^2}{\Delta'_r(r_0)^2} + 8 r_0 \frac{\Delta_r(r_0)}{\Delta'_r(r_0)^2} [2 r_0 \Delta_y(y_\mathcal{O}) + \Sigma(r_0, y_\mathcal{O}) \Delta_r'(r_0)] - \Sigma^2(r_0, y_\mathcal{O}) \right]^{1/2} \,.
	\end{align}
	\end{subequations}
	With these expressions, one has access to the positions of the impact of the nearly critical photons building the critical curve. Notice that at this stage, the couple $(\alpha_c, \beta_c)$ depends on the parameters of the black hole, the position of the observer's screen $(r_{\cO}, y_{\cO})$ which is fixed, and finally on the position $r_0$ at which the photon is emitted within the photon shell, i.e. the region spanned by spherical photon orbits. Contrary to the other parameters, one has $r_0 \in [r^{-}_{\text{ph}}, r^{+}_{\text{ph}}]$ such that this parameter is not fixed. Therefore, for now on, we write the couple of Cartesian coordinates as $\alpha_c(r_0)$ and $\beta_c(r_0)$ to make the dependency on this parameter explicit.
	
	It is useful to introduce alternative coordinates on the screen of the observer. Let us decompose the dimensionless Cartesian coordinates $(\alpha(r_0), \beta(r_0))$ into polar ones as
	\be
	\alpha(r_0) := \cR(r_0) \cos{\phi(r_0)} \,, \qquad \beta (r_0)= \cR(r_0) \sin{\phi(r_0)} \,,
	\ee
	where $\cR(r_0)$ is a dimensionless radius and $\phi(r_0)$ the angle w.r.t the $\alpha$-axis. Notice that here, $r_0$ plays the role of a curvilinear abscissa for the critical curve. Let us introduce a basis of orthogonal unit vectors on the plane $(\alpha, \beta)$ denoted $\left( \vec{e}_{\alpha}, \vec{e}_{\beta}\right)$. The position on the curve is given by
	\be
	\vec{r}_c =  \cR_c(r_0) \vec{u} \qquad \text{with} \qquad \vec{u} = \cos{\phi_c(r_0)} \vec{e}_{\alpha} +  \sin{\phi_c(r_0)} \vec{e}_{\beta} \,.
	\ee
	A direct computation shows that the coordinates $(\cR_c(r_0), \phi_c(r_0))$ are related to the black hole parameters by
	\begin{subequations}
	\begin{align}
	\cR_c(r_0) & = \sqrt{\alpha_c^2(r_0)  + \beta_c^2(r_0) } = \frac{1}{r_\mathcal{O}} \qty[\Delta_y(y_\mathcal{O}) + 2\Sigma(r_0, y_\mathcal{O}) + 8[2r_0 - \Delta_r'(r_0)] \frac{\Delta_r(r_0)}{\Delta_r'(r_0)^2}]^{1/2} \,,\\
	\text{tan}[\phi_c(r_0) ] & = \frac{\beta_c(r_0) }{\alpha_c(r_0) } = \mp \frac{\sqrt{ 16 r_0^2 \Delta_r(r_0)\Delta_y(y_\mathcal{O}) - \big[\Sigma(r_0, y_\mathcal{O}) \Delta_r'(r_0) - 4r_0 \Delta_r(r_0)\big]^2 }}{  \Delta'_r(r_0) \left( \Sigma(r_0, y_\mathcal{O}) + \Delta_y (y_{\cO})\right) -4r_0\Delta_r(r_0)} \,.
	\end{align}
	\end{subequations}
	These relations are not yet the ones relevant to interprete the interferometry signal. As discussed in detail in \cite{Gralla:2020nwp, Gralla:2020yvo}, the relevant information is contained in the angle $\varphi(r_0)$ which encodes the ratio between the rate of change of the curve w.r.t the $\alpha$-axis and the rate of change of the curve w.r.t the $\beta$-axis. Concretely, this angle is given by
	\begin{align}
	\label{eq:coord-critical-curve}
	\boxed{\text{tan}[\varphi_c(r_0) ]  = - \frac{ \partial_{r_0}\alpha_c(r_0) }{\partial_{r_0}\beta_c(r_0) } = \pm \frac{\sqrt{16 r_0^2 \Delta_r(r_0)\Delta_y(y_\mathcal{O}) - \big[\Delta_r'(r_0) \Sigma(r_0, y_\mathcal{O}) - 4 r_0 \Delta_r(r_0)\big]^2}}{4 r_0 (\Delta_r(r_0) - \Delta_y(y_\mathcal{O})) - \Sigma(r_0, y_\mathcal{O}) \Delta_r'(r_0)} \,.}
	\end{align}
 This is the key relation introduced in \cite{Gralla:2020yvo} to reconstruct the image of the critical curve.
 
 This concludes the derivation of the critical curve and its properties for the general KOS family. We now use these general results to discuss the status of the test of the Kerr hypothesis using the shape of the critical curve. 

\newpage

	\section{Parameter estimations beyond Kerr: concrete examples}
	\label{sec:examples}
			In this section, we use the previous results to address two key points. 
			
			First, we demonstrate that the formula obtained in the previous section allows us to efficiently capture the position and shape of the photon ring for different examples of Kerr-like objects belonging to the KOS family. Two of these examples are known models but the remaining two exhibit new types of corrections which have not appeared in the literature so far, illustrating the power of the new parametrization.  Contrary to most investigation which only focus on the position of the equatorial radius of the photon ring, we derive the positions of the horizon, the radial and polar positions of the photon ring and the critical parameters $(\ell_c, k_c)$ for spherical photon orbits for each examples. 
			
			Second, we use these concrete examples to discuss to what extend the shape of the photon ring can provide a sharp test of the Kerr hypothesis, and therefore of GR. In~\cite{Gralla:2020yvo}, it was proposed to fit the critical curve (\ref{eq:coord-critical-curve}), denoted $f(\varphi)$, by a \textit{phoval}, described by
	\begin{equation}
		f(\phi) = R_0 + \sqrt{R_1^2 \sin^2\phi + R_2^2 \cos^2\phi} + (X - \chi) \cos\phi + \arcsin(\chi \cos\phi) \,,
		\label{eq:phoval}
	\end{equation}
	where $R_0$, $R_1$, $R_2$, $X$ and $\chi$ are scalar parameters. The link of each of these parameters with $M$, $a$ and the inclination $y_\mathcal{O}$ can only be found numerically. They key point is that this \textit{phoval} (or circplise) function fit to a great accuracy the Kerr critical curve. In the following, we confront this phoval fitting function introduced in \cite{Gralla:2020yvo} to the four examples and study the degeneracy in the estimation of the black hole parameters.  
		
	\subsection{The Kerr black hole}
	\label{sec:kerr-photon-ring}
	
	As a first test of our general close formula, let us treat the special case of the Kerr black hole. Consider again the following change of coordinates
	\begin{equation}
	\tag{\ref{relcoord}}
		y = a \cos{\theta} \,, \qquad \varphi = \frac{\phi}{a} \,, \qquad \tau = t - a \phi\,,
	\end{equation}
	which relates the coordinates $(\tau, r, y, \varphi)$ for the Boyer-Lindquist coordinates $(t, r, \theta, \phi)$. Then, the Kerr metric is recovered by choosing
	%
	\begin{equation}
	\tag{\ref{KerrFunc}}
		\Delta^{\text{Kerr}}_r =  r^2 + a^2 - 2 Mr \,, \qquad \Delta^{\text{Kerr}}_y = a^2 - y^2 = a^2 \sin^2{\theta} \,.
	\end{equation}
	The Kerr outer horizon is located at 
	\be
	r_h = M + \sqrt{M^2-a^2} \,.
	\ee
	The parameters $(\ell_c, k_c)$ for the spherical photon orbits of Kerr satisfy (\ref{sol}) which reads
	\begin{align}
	k_c = \frac{4r^2_0 (r^2_0 - 2 Mr_0 + a^2)}{(r_0 - M)^2} \,, \qquad \ell_c = - \frac{r_0(r_0^2 -3M r_0 + 2 a^2)}{r_0 -M} \,,
	\end{align}
	where $r_0$ is a radius satisfying $r_0 > M$ and belonging to the photon ring. The range of $r_0$ is constrained by the inequality~\eqref{ph}. Plugging the Kerr functions (\ref{KerrFunc}), this equation reduces to the standard cubic polynomial equation
	\begin{equation}
		r^3 - 6M r^2  + 9 M^2 r - 4 a^2 M \leqslant  0\,,
	\end{equation}
	which determines the minimal and maximal equatorial radii of the Kerr photon ring given by
		\begin{equation}
		r_\mathrm{ph}^\pm = 2M \Big[1 + \cos(\frac23 \xi^\pm)\Big]  \qq{with} \xi^\pm = \arccos(\pm\frac{a}{M}) \,.
		\label{eq:photonring-kerr}
	\end{equation} 
	such that $r_0 \in [r_\mathrm{ph}^{+}, r_\mathrm{ph}^{-}]$. 
	As we have seen, the polar motion depends on the value of the couple $(\ell, k)$. Besides the equilibrium positions $y^{\pm}_e = \{ 0, \pm\ell\}$, one has also other positions described in  Section~(\ref{polarmo}). 
	
\begin{figure}[!htb]
		\centering
		\includegraphics{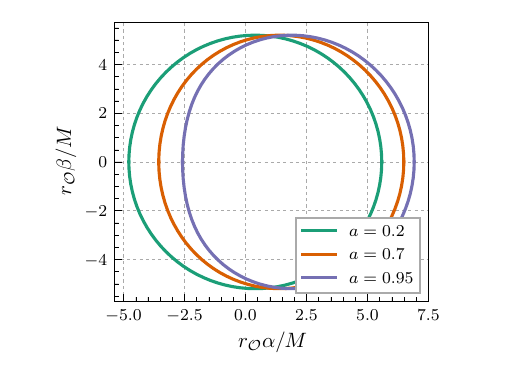}
		\caption{Critical curve of the Kerr black hole for $M=1$, $y_\mathcal{O}=0$ and various values of $a$.}
		\label{fig:shadow-Kerr}
	\end{figure}

	Therefore, the closed formula obtained in the previous section consistently reproduce the known results for the positions of the photon ring and the parameters of the spherical photon orbits for the Kerr black hole. Using the above results, we can plot the parametric form of the critical curve (\ref{eq:coord-critical-curve}) for different values of the spin $a$. It is shown in fig.~\ref{fig:shadow-Kerr}.
	
	Having tested our formula on the standard Kerr black hole example, our goal now is to consider different examples of Kerr-like objects and reproduce the very same study for each of them. However, before presenting these Kerr-like objects, let us already make an important remark. It is standard to work with dimensionless screen coordinates $(r_{\cO}\alpha/r_{\ast}, r_{\cO}\beta/r_{\ast})$ where $r_{\cO}$ is the radial position of the observer and $r_{\ast}$ s usually taken to be the mass of the black hole (see fig~\ref{fig:shadow-Kerr}). Therefore, the mass of the black hole will directly affect the overall scaling of the image. However, while the mass of the Kerr black hole admits a well-defined interpretation as an asymptotic charge of the geometry, this is no longer true in modified gravity. Indeed the standard Komar charge one uses in GR is generically modified when one considers alternatives (and in particular higher order) lagrangians. While the modifications to the Komar charge can be computed in principle using the covariant phase space method, this point has not been studied systematically such that given an ad hoc Kerr-like geometry, there is no way to know what one should interprete what is the quasi-local asymptotic mass of the object as this notion depends on the theory which admits this geometry as a solution. In order to bypass this difficulty when dealing with ad hoc models for Kerr-like object, one way is to use the position of the horizon of the geometry which can be extracted solely form the metric in order to rescale the screen coordinates. This is the choice we adopt in the following when plotting the shadow of Kerr like objects.
	
	\subsection{Radial deformations to Kerr}

In this section, we consider three examples of deviations to Kerr which are purely radial, i.e. they correspond to different choices of $\Delta_r(r)$. The first two are known examples of Kerr-like objects that have been studied in the literature: the Kerr-MOG black hole derived in \cite{Moffat:2014aja} and the regular rotating black hole introduced by Simpson and Visser in \cite{Simpson:2021dyo}. The third example is a new type of correction that does not appear elsewhere.
	
		\subsubsection{The Kerr-MOG black hole}
		
		Consider first The Kerr-MOG black hole discussed in \cite{Moffat:2014aja}. This rotating black hole geometry is a solution of the so called Scalar-Vector-Tensor (SVT) theory introduced in \cite{Moffat:2005si}.  Its shadow was previously studied in \cite{Liu:2024lbi}. The geometry is characterized by a new parameter $\alpha$ which can be understood as a coupling constant of the SVT lagrangian which changes the Newton constant $G_N$ to effective Newton constant of the form $G = (1+\alpha)G_{N}$. It can be shown that the new Kerr-like solution of this theory enjoys a modified radial function $\Delta_r(r)$ given by:
		\be
		\Delta_r(r) = r^2 - 2GMr + a^2 + \alpha G_N G M^2 = \Delta^{\text{Kerr}} + \alpha G_N G M^2 \,,
		\ee
		where $G = G_N (1+\alpha) = 1+\alpha$ when $G_N = 1$. The outer horizon is located at 
		\be
		r_h = (1+\alpha) M \left(1 + \sqrt{1 - \frac{a^2}{(1+\alpha)^2 M^2}- \frac{\alpha}{1+\alpha}} \right)
		\ee
		By construction, this Kerr-like geometry belongs to the KOS family and therefore, one can use the previous results to obtain analytical expressions for the different locus of points of its photon rings and the parameters of its spherical photon orbits.
		The couple $(\ell_c, k_c)$ is given by
		\begin{subequations}
		\begin{align}
		\label{sol2}
		\ell_c &= -\frac{r \left(2 a^2+2 \alpha  (\alpha +1) M^2-3 (\alpha +1) M r+r^2\right)}{r-(\alpha +1) M} \,,\\ k_c &= \frac{4 r^2 \left(a^2+\alpha  (\alpha +1) M^2-2 (\alpha +1) M r+r^2\right)}{(\alpha  M+M-r)^2} \,.
		\end{align}
		\end{subequations}
		The parameter $r_0$ ranges between the minimal and maximal positions (written at first order in $\alpha$ for clarity here)
		\be
		r_\mathrm{ph}^\pm = 2M \Big[1 + \cos(\frac23 \xi^\pm)\Big] + \frac13 \alpha M \qty[7 + \frac{2a^2 \cos(\frac23\xi^\pm)}{a^2 + M^2 (\cos(\frac23 \xi^\pm) + \cos(\frac43 \xi^\pm)}] + \mathcal{O}(\alpha^2)  \,,
		\ee
		where $\xi^\pm$ is the quantity defined for Kerr in~\eqref{eq:photonring-kerr}.
		Notice that since the geometry does not have any polar deformation, i.e. $\Delta_y (y) =\Delta^{\text{Kerr}}_y (y) $, the expressions for the equilibrium polar positions $y^{\pm}_e$ have the same expressions as for Kerr. 
		
\begin{figure}[!htb]
	    \captionsetup[subfigure]{justification=Centering}
		\centering
		\begin{subfigure}[t]{0.45\textwidth}
            \centering
            \includegraphics{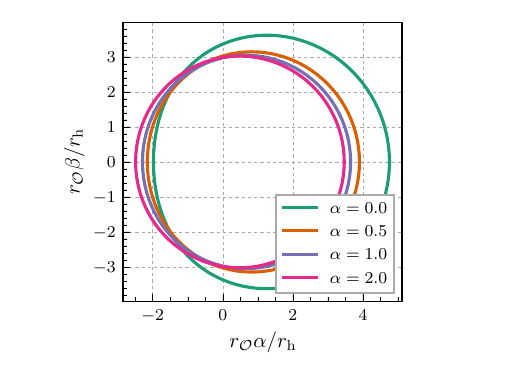}
            \caption{$y_\mathcal{O} = 0$}
        \end{subfigure}\hspace{0cm}%
        \begin{subfigure}[t]{0.45\textwidth}
            \centering
            \includegraphics{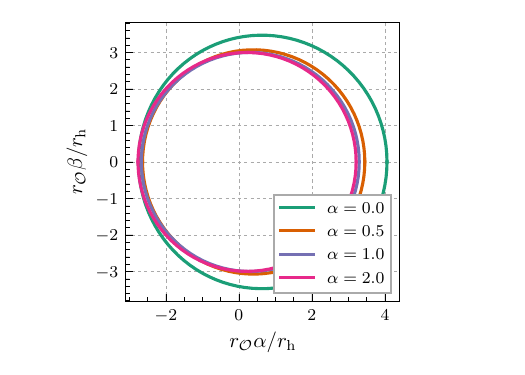}
            \caption{$y_\mathcal{O} = 0.8$}
        \end{subfigure}
		\caption{Critical curve of the Kerr-MOG black hole for $M=1$, $a = 0.9$ and different values of $y_\mathcal{O}$.}
		\label{fig:shadow-MOG}
	\end{figure}

		With these results, one can use the general formula \eqref{eq:coord-critical-curve} to plot the critical curve of the Kerr-MOG black hole for i) four different value of the new parameter $\alpha$ and ii) two different values of the inclination $y_{\cO}$. It is represented in Fig~\ref{fig:shadow-MOG}.
		
	As a first remark, one can see that the deviations w.r.t the Kerr shadow (depicted in green) are qualitatively more important at high inclination, i.e. for $y_{\cO}=0$, in agreement with previous studies \cite{Staelens:2023jgr}. However, the shape of the critical curve represented here does not provide in itself a precise way to compare the Kerr and Kerr-MOG black holes. To confront the two models, one needs to perform a parameter estimations of the black holes parameters. 
	
	To that end, one can use the fitting function introduced by Gralla and Lupsasca and which consists in the phoval (or circlipse) defined by the equation (\ref{eq:phoval}) to test the two models. This is summarized in the Table~\ref{tab:degeneracy-circlipse}. As a result, we find that this fitting function admits a non-trivial degeneracy in the estimation of the parameters $(R_0, R_1, R_2)$. Concretely, for this critical curve, the circlipse fitting test cannot distinguish between a Kerr black hole geometry with mass and spin $(M,a) =(1, 0.2)$ and a Kerr-MOG black hole with parameters $(M,a)=(0.922, 1.923)$ and $\alpha =0.101$. This example provides the first explicit and concrete proof that the circlipse test does not, by itself, provide a sharp test of the Kerr hypothesis. It suggests that in order to properly test the Kerr hypothesis, one has to be able to test the relation between the mass and angular momentum which characterize a given geometry. Obtaining independent measures of the mass and the spin of the black holes appear therefore as a mandatory task to break the degeneracy presented here. We stress that this example (and the analytical derivation of the critical curve used here) could not have been obtained without the use of the KOS parametrization introduced in this work. Let us now turn to other examples of Kerr-like objects.
	
		\begin{table}[!htb]
	    \centering
	    \begin{tabular}{|c|c|c|c|c||c|c|c|c|}
	        \hline
	         Parameters & $M$ & $a$ & $\alpha$ & $y_\mathcal{O}$ & $R_0$ & $R_1$ & $R_2$ & Residuals\\\hline
	         Kerr & \num{1} & \num{0.2}& \diagbox[innerwidth=1em, height=\line]{}{} & 0 & \num{10.0} & \num{0.394} & \num{0.370}& \num{1.88e-5}\\\hline
	         MOG & \num{0.922} & \num{0.193} & \num{0.101} & 0 & \num{10.0} & \num{0.395} & \num{0.371} & \num{1.86e-5} \\\hline
	         \end{tabular}
	    \caption{Example of degeneracy between the Kerr black hole and the Kerr-MOG model.}
	    \label{tab:degeneracy-circlipse}
	\end{table}
		
		\subsubsection{The regular Simpson-Visser rotating model}
		
		 We now consider a model of rotating regular black hole. Such models have been studied intensively in the literature \cite{Carballo-Rubio:2018jzw, Lan:2023cvz, Torres:2022twv}. To a large extent, the regularization of the interior singularity of spherically symmetric black holes have been performed by introducing an effective de Sitter and thus repulsive core. Yet, other type of regularization have been studied and it has been shown that one could obtain interesting ad hoc regular black hole geometries by replace the deep interior geometry by a Minkowski core \cite{Simpson:2019mud}. This approach was extended to rotating black hole and an explicit regular Kerr-like black hole dubbed the "Eyes of the Storm" (EOS) was presented in \cite{Simpson:2021dyo}. The modification amounts at the a change of the mass which becomes $M(r) = M e^{- \ell M /r}$ where $\ell$ is the new UV cut-off encoding the scale at which the effective quantum effects are triggered. For a given $(M, a)$, $\ell=0$ yields the Kerr black hole. Furthermore, possible values of $\ell$ are bounded from above: $0 \leq \ell \leq l_\text{extr}$. Image of this regular black hole have already been studied in \cite{Benavides-Gallego:2024hck}
		 
\begin{figure}[!htb]
	    \captionsetup[subfigure]{justification=Centering}
		\centering
		\begin{subfigure}[t]{0.45\textwidth}
            \centering
            \includegraphics{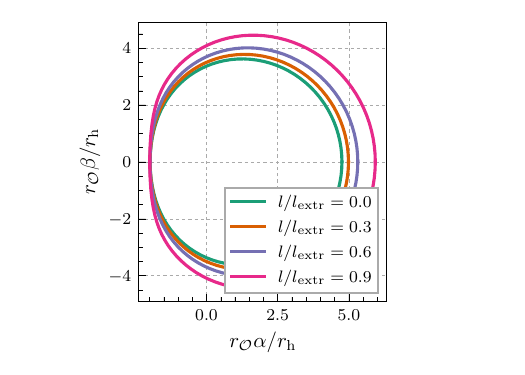}
            \caption{$y_\mathcal{O} = 0$}
        \end{subfigure}\hspace{0cm}%
        \begin{subfigure}[t]{0.45\textwidth}
            \centering
            \includegraphics{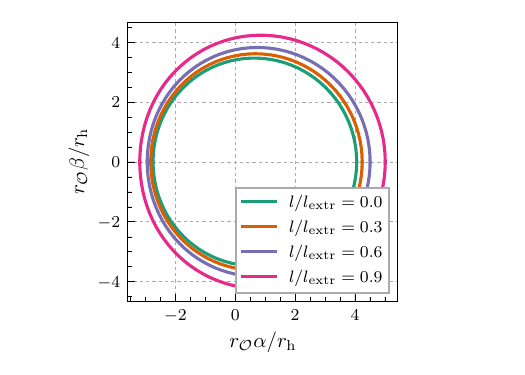}
            \caption{$y_\mathcal{O} = 0.8$}
        \end{subfigure}
		\caption{Critical curve of the Simpson-Visser black hole for $M=1$, $a = 0.9$ and different values of $y_\mathcal{O}$.}
		\label{fig:shadow-SV}
	\end{figure}

		 The modification amounts at a change of $\Delta(r)$ of the form
\be
\Delta_{\text{EOS}} = r^2 + a^2 - 2 r M(r) = r^2 + a^2 - 2r M e^{- \ell M/r} \,,
\ee
belonging thus to the KOS family. Unfortunately, the choice of the exponential function in the function $M(r)$ does not allows one to find closed analytical formulae for the positions of the horizons. At best, one can provide an expression for the correction to the Kerr horizon when $\ell$ is considered small, which gives 
\be
r^{\text{EOS}}_h = M + \sqrt{M^2-a^2 - 2M\ell + \cO(\ell^2)} < r^{\text{Kerr}}_h
\ee
Using the previous relations, we now derive the parameters $(\ell_c, k_c)$ for the critical spherical photon orbits which read
\begin{align}
\ell_c &= \frac{r^2 \left(M (3 r-M \ell )-\left(2 a^2+r^2\right) e^{\frac{M \ell }{r}}\right)}{r^2 e^{\frac{M \ell }{r}}-M (M \ell +r)} \\
k_c & = \frac{16 r^2 \left(a^2+r \left(r-2 M e^{-\frac{M \ell }{r}}\right)\right)}{\left(2 r-\frac{2 M e^{-\frac{M \ell }{r}} (M \ell
   +r)}{r}\right)^2}
\end{align}
where the position of the photon ring is given (at first order in $\ell$ for clarity) by
\begin{align}
r^{\pm}_{\text{ph}} & = 2M \Big[1 + \cos(\frac23 \xi^\pm)\Big] + \frac23 \ell M \qty[-2 + \frac{a^2}{a^2 + M^2 (\cos(\frac23 \xi^\pm) + \cos(\frac43 \xi^\pm)}] + \mathcal{O}(\ell^2)  \,,
		\end{align}
		where $\xi^\pm$ is the quantity defined for Kerr in~\eqref{eq:photonring-kerr}.
The critical curve of the EOS black hole for a few different values of the parameter $\ell$ and at two different inclinations are shown in fig.~\ref{fig:shadow-SV}. While the green curve represents the Kerr black hole critical curve, we vary the parameter $\ell$ up to its nearly maximal value corresponding to the extremality. 
	Following the same procedure as above, we also find that the circplispe fitting function suffers from a degeneracy as shown by the concrete example of parameters estimation presented in Table~\ref{tab:degeneracy-circlipse2}. This second example provides thus a confirmation of the result obtained for the study of the Kerr-MOG model. We can now turn to yet new examples of Kerr-like objects which have not been treated so far. 
		\begin{table}[!htb]
	    \centering
	    \begin{tabular}{|c|c|c|c|c||c|c|c|c|}
	        \hline
	         Parameters & $M$ & $a$ & $\ell$ & $y_\mathcal{O}$ & $R_0$ & $R_1$ & $R_2$ & Residuals\\\hline
	         Kerr & \num{1} & \num{0.2}& \diagbox[innerwidth=1em, height=\line]{}{} & 0 & \num{10.0} & \num{0.394} & \num{0.370}& \num{1.88e-5}\\\hline
	         SV & \num{1.32} & \num{0.1} & \num{0.82} & 0 & \num{10.0} & \num{0.402} & \num{0.380} & \num{1.53e-5} \\\hline
	    \end{tabular}
	    \caption{Example of degeneracy between the Kerr black hole and the Simpson-Visser model. The observer is at $y_\mathcal{O} = 0$ in both cases.}
	    \label{tab:degeneracy-circlipse2}
	\end{table}
	
	\subsubsection{Log corrections}
	\label{sec:log-corrections}
	
\begin{figure}[!htb]
	    \captionsetup[subfigure]{justification=Centering}
		\centering
		\begin{subfigure}[t]{0.45\textwidth}
            \centering
            \includegraphics{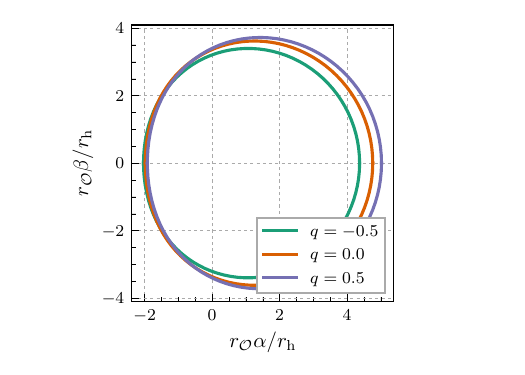}
            \caption{$y_\mathcal{O} = 0$}
        \end{subfigure}\hspace{0cm}%
        \begin{subfigure}[t]{0.45\textwidth}
            \centering
            \includegraphics{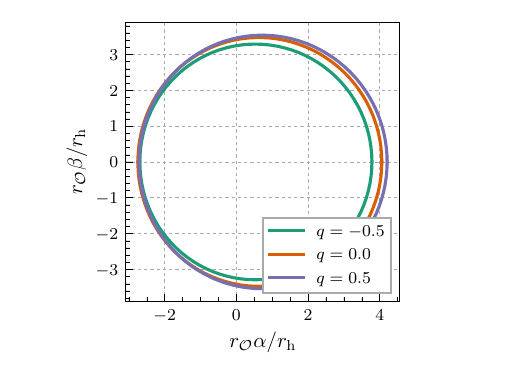}
            \caption{$y_\mathcal{O} = 0.8$}
        \end{subfigure}
		\caption{Critical curve of the log-corrected black hole for $M=1$, $a = 0.9$ and different values of $y_\mathcal{O}$.}
		\label{fig:shadow-log}
	\end{figure}

	An interesting outcome of the KOS parametrization used in this work is to allows us to consider new ad hoc modification to the Kerr geometry which will not break the Killing tower. As a concrete illustrative example, we now consider a one-parameter family of deviations from Kerr defined by
	\begin{equation}
		 \Delta_r = r^2 - 2M r + a^2 + q M^2 \log(\frac{r}{M}) \,,
	\end{equation}
	This log-modification still preserves the asymptotical flatness condition as $\Sigma \sim r^2$ at large $r$. However, to our knowledge, it has not been considered in the literature. We write analytical formulas at first order in the dimensionless variable $q$ for clarity; however, all results are exact in $q$, in particular the plots shown on fig~\ref{fig:shadow-log}. The horizon $r_\mathrm{h}$ of the Kerr black hole is changed at first order as follows:
	\begin{equation}
		r_\mathrm{h} = M + \sqrt{M^2 - a^2} - q \frac{M}{2} \frac{\log(1 + \sqrt{1 - (a/M)^2})}{\sqrt{1 - (a/M)^2}} + \mathcal{O}(q^2) \,.
	\end{equation}
	A direct computation shows that the couple $(\ell_c, k_c)$ is given by
		\begin{align}
		\label{sol2}
		\ell_c &= r \left(r-\frac{4 r \left(a^2+M^2 q \log \left(\frac{r}{M}\right)+r (r-2 M)\right)}{M^2 q-2 M r+2 r^2}\right) \,,\\
		k_c &= \frac{16 r^4 \left(a^2+M^2 q \log \left(\frac{r}{M}\right)+r (r-2 M)\right)}{\left(M^2 q-2 M r+2 r^2\right)^2} \,.
		\end{align}
	The parameter $r_0$ ranges between the minimal and maximal positions given by
	\begin{equation}
		r_\mathrm{ph}^\pm = 2M \Big[1 + \cos(\frac23 \xi^\pm)\Big] + q d_\pm^{(1)} + \mathcal{O}(q^2)\,, \quad \xi^\pm = \arccos(\pm\frac{a}{M}) \,.
	\end{equation}
	The corrections $d_\pm^{(1)}$ are such that
	\begin{multline}
		d_\pm^{(1)} =  - \frac{M}{2} \frac{ \left( 4 \log[4 \cos^2(\xi^\pm/3)] - 1 \right) \left( \cos(4\xi^\pm/3) + \cos(2\xi^\pm) \right) + \left( 2 \log[4 \cos^2(\xi^\pm/3)] - 1 \right)  (a/M)^2 }{2 + 2\cos(2\xi^\pm) + 3\cos(4\xi^\pm/3) + 3 \cos(2\xi^\pm/3) - (a/M)^2 }\,.
	\end{multline}
	This first example shows that one can use the symmetries of the Kerr off shell system to analytically capture the modifications of the photon ring for some new deviations w.r.t. Kerr.
	Using the results of~\ref{sec:shadow}, we can numerically trace the shadow of the black hole. The result is shown on fig~\ref{fig:shadow-log}.
	
	\subsection{Polar deformations to Kerr}
	
	\label{polardef}
	
\begin{figure}[!htb]
	    \captionsetup[subfigure]{justification=Centering}
		\centering
		\begin{subfigure}[t]{0.45\textwidth}
            \centering
            \includegraphics{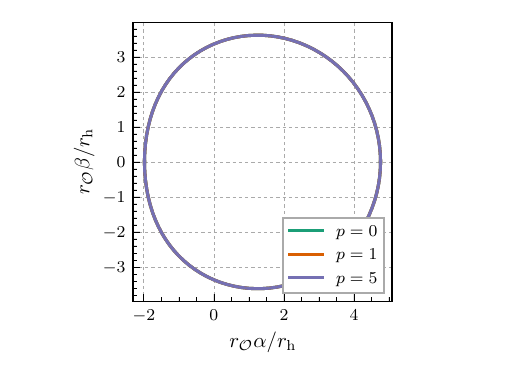}
            \caption{$y_\mathcal{O} = 0$}
        \end{subfigure}\hspace{0cm}%
        \begin{subfigure}[t]{0.45\textwidth}
            \centering
            \includegraphics{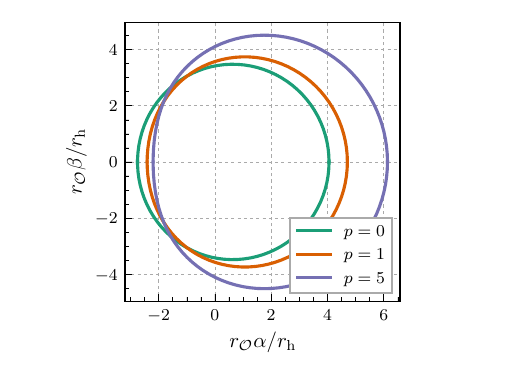}
            \caption{$y_\mathcal{O} = 0.8$}
        \end{subfigure}
		\caption{Critical curve of the polar-corrected black hole for $M=1$, $a = 0.9$ and different values of $y_\mathcal{O}$. When $y_\mathcal{O} = 0$, the black hole is seen from its equatorial plane, and the polar corrections do not come into play: the critical curves for all values of $p$ are the same.}
		\label{fig:shadow-polar}
	\end{figure}

	We now turn to a situation in which $\Delta_y$ deviates from the Kerr solution while $\Delta_r$ is unchanged. In order to maintain the reflection symmetry, we take the parametrization
	\begin{equation}
		\Delta_y = a^2 - y^2 + p y^4 \,.
	\end{equation}
	Similarly to sec~\ref{sec:log-corrections}, we present results at first order in $p$ for clarity but our results are exact in this variable. The horizon $r_\mathrm{h}$ is unchanged since $\Delta_r$ is the same, as well as $\ell_c$, $k_c$ and $r^{\pm}_{\text{ph}}$ which are the same as in the Kerr case (see section~\ref{sec:kerr-photon-ring}). However the value of $y_\mathrm{m}$, which corresponds to the maximal value of $y$ for a null geodesic with $\ell > 0$, is modified: one has
	\begin{equation}
		y_\mathrm{m} = \sqrt{\frac12 \Big( -(k + 2\ell) + \sqrt{k}\sqrt{k + 4 (a^2 + \ell)} \Big)} + q y_\mathrm{m}^{(1)} + \mathcal{O}(q^2) \,,
	\end{equation}
	with
	\begin{equation}
		y_\mathrm{m}^{(1)} = - \frac{\big(k-\sqrt{k} \sqrt{4 a^2+k+4 \ell }\big) \big(\sqrt{k} \sqrt{4 a^2+k+4 \ell }-k-2 \ell\big)^{3/2}}{4 \sqrt{2} a^2 \big(-\sqrt{k} \sqrt{4 a^2+k+4 \ell }+4 a^2+k+4 \ell \big)} \,.
	\end{equation}
	Using the results of section~\ref{sec:shadow}, we can numerically trace the shadow of the black hole. The result is shown on fig~\ref{fig:shadow-polar}.

	\newpage
	
	\section{Discussion}

	In this work, we have shown how the Kerr off shell (KOS) family of geometries introduced in \cite{Krtous:2006qy, Kubiznak:2006kt, Krtous:2008tb} allows one to efficiently parametrize beyond Kerr geometries. Using this new parametrization, we have studied the deformations of the shape of the black hole photon ring image and confronted the ability to distinguish between Kerr and beyond Kerr geometries using the circlipse fitting function proposed by Gralla and Lupsasca. Let us summarize the main points of our study. \bigskip
	
 \textit{The KOS parametrization:} The KOS family of metrics stands as the most general set of geometries preserving the Killing tower. This structure uncovered in \cite{Krtous:2006qy} encodes the fundamental explicit and hidden symmetries of the Kerr geometry. In particular, any KOS representant is of Petrov type D and is equipped with a rank-$2$ closed conformal Killing-Yano tensor and its dual. In turn, this ensures that the geometry also admits a Killing tensor which makes the geodesic motion integrable. Remarkably, despite preserving these hidden symmetries, the KOS parametrization still allows for a large set of deformations to the Kerr geometry encoded in the two free functions $\Delta_r(r)$ and $\Delta_y(y)$. On one hand, $\Delta_r(r)$ allows for general radial deformations and coincides with the restricted class I family of beyond Kerr geometries studied in \cite{Yagi:2023eap} where $\Delta_r(r) = A_5(r)$. This radial deformation allows one to reproduce a large class of \textit{ad hoc} models studied so far in the literature. On the other hand, the KOS parametrization provides another type of polar deformations encoded in $\Delta_y(y)$ which have not been studied so far. By construction, such polar deformations preserve the symmetries of Kerr, allowing for the analytic exploration of new Kerr-like objects such as the example studied in Section~\ref{polardef}. \bigskip \\
\textit{Analytic characterization of the photon shell:} Since the image of the black hole photon ring is generated by the geodesics which are nearly critical, i.e. geodesics which are infinitely close to the unstable critical spherical photon orbits,  we have provided a detailed study of this family of geodesics. One of the remarkable outcomes of the KOS family is to allow us to obtain an analytical characterization without fixing the free functions of the metric. Concretely, for an arbitrary KOS geometry, the radial and polar motions are determined by the two potentials (\ref{potgeo11} -- \ref{potgeo22}) which depend respectively on the radial and polar deformations $\Delta_r(r)$ and $\Delta_y(y)$. The critical spherical photon orbits which satisfy (\ref{crit}) are characterized by a reduced angular momentum and Carter constant $(\ell_c, k_c)$ given by (\ref{sol}). These expressions depend solely on the function $\Delta_r(r)$ and the radial equilibrium position $r_0$ which can be obtained only once a given choice of $\Delta_r(r)$ is made. The extent of the region in which the spherical orbits are allowed, i.e. the photon shell, is constrained by the inequality (\ref{ph}). For all the examples studied in this work, the allowed position $r_0$ is restricted to a range $[r_{\text{ph}}^{-}, r_{\text{ph}}^{+}]$ with $r_{\text{ph}}^{-} > r_\mathrm{h}$ where $r_\mathrm{h} >0$ stands for the position of the outer horizon. One the other hand, we have provided a detailed classification of the possible polar equilibrium positions for such critical spherical photon orbits. Therefore, for a given KOS representant, the above formula allows one to fully characterize the photon shell geometry. \bigskip \\
 \textit{Non-spherical unstable critical photon orbits:} Notice that a Kerr-like geometry could in principle allow for non-spherical orbits within the photon shell, i.e. photon orbits with $ \cV'_r(r) \neq 0$ which admit radial turning points $r_{\ast}$ (at which $ \cV_r(r_{\ast}) = 0$). The existence of such orbits depends on the form of the function $\Delta_r(r)$. For the Kerr geometry, one can show that such non-spherical critical orbits do not exist. For the general KOS geometry, we have not been able to provide a definite proof that they are forbidden. However, for all the examples considered here, one can show that such non-spherical critical orbit are not allowed. \bigskip \\
 \textit{Analytic parametrization for the critical curve:}  The main outcome of this study is the analytical derivation of the critical curve for an arbitrary KOS geometry given by Eq~(\ref{eq:coord-critical-curve}). This result provides a ready-to-use parametrization of the critical curve of beyond Kerr geometries which depends explicitly on the two free functions $\Delta_r(r)$ and $\Delta_y(y)$. To our knowledge, such analytic expressions has not been obtained yet for such a general class of rotating compact objects. At this stage, it is important to underline that the critical curve is not an observable \textit{per se}. The interest in having an analytic prediction for this object lies in that fact that the photon subrings converge quickly towards this theoretical curve, making it the key object to parametrize. Let us emphasize that such an analytic result is made possible by to the integrability of the geodesic motion and thus by the preservation of the Kerr symmetries. \bigskip \\
 \textit{The circlipse fitting function and its degeneracy:} Equipped with this analytical parametrization of the critical curve, we have confronted the circlipse fitting functions on four different examples of Kerr-like geometries. We show that at maximal inclination, i.e. $y_{\cO} =0 $, an accurate fit of the Kerr black hole for a given $(M, a)$ can always be degenerate with another equally accurate fit for the Kerr-like object for another set of parameters $(M, a, \alpha)$ where $\alpha$ is the new parameter encoding the deviation to Kerr. Therefore, alone, the circlipse does not allow for a test the Kerr hypothesis, confirming recent results \cite{Staelens:2023jgr}. In order to break this degeneracy, an independent measurement of the mass and spin of the compact object is needed (see~\cite{Grould:2017bsw,Bambi:2020jpe}). Notice that this degeneracy between different models of compact objects add up to other type of degeneracy, in particular degeneracy coming from the astrophysical modeling of the disk accretion \cite{Paugnat:2022qzy}. It follows that in order to use black hole imaging as a sharp test of GR, important challenges are ahead of us to systematically alleviate the different degeneracies.  \bigskip \\
 \textit{Kinematical versus dynamical tests:} Another interesting outcome of working out the critical curve and the resulting image of beyond Kerr geometries is to underline some conceptual difficulties when dealing with the notion of mass and spin. Indeed, for the Kerr black hole which solve the Einstein equations, the notions of mass and spin are well-understood in term of asymptotic Komar charges whose expressions are derived from the Einstein-Hilbert action. However, when considering an \textit{ad hoc} geometry which is not solution of any given theory of gravity, such as the Simpson-Visser model studied here, we have no recipe to identify a well defined notion of quasi-local mass or angular momentum. For this reason, while considering \textit{ad hoc} geometries provide a useful first step to study beyond Kerr phenomenology, it is mandatory to associate such models to given theories to interpret the different parameters of the geometry as well-defined asymptotic quasi-local charges \footnote{See \cite{Minamitsuji:2023nvh} for a study of quasi-local energy using the covariant phase space in Horndeski}. This task represents another challenge in order to use black hole imaging to test alternatives theories of gravity such as scalar-tensor theories.  (See \cite{BenAchour:2024hbg} for review of rotating black holes and possible solution generating method in such modified theories).  An interesting example treated here is the Kerr-MOG black hole, which is a \textit{bona fide} solution of a scalar-vector-tensor theory where the additional parameter $\alpha$ plays the role of a coupling constant in the lagrangian of the theory. Thus, constraining the parameter $\alpha$ allows one to confront two theories and not only two geometries. From this perspective, the natural next step will be to identify suitable family of theories admitting the KOS geometries as exact solutions. This will be presented in the near future.

\subsection*{Acknowledgments}

We thank Frédéric Vincent for useful discussions. 
EG acknowledges funding by l’Agence Nationale de la Recherche, projects StronG ANR-22-CE31-0015-01 and Einstein PPF ANR-23-CE40-0010-02. JBA and HR thank the Perimeter Institute for Theoretical Physics for its hospitality at the beginning of this project.

	\newpage

	\bibliographystyle{JHEP}
	\bibliography{biblio}
	
\end{document}